\pdfoutput=1
\RequirePackage{ifpdf}
\ifpdf 
\documentclass[pdftex]{sigma}
\else
\documentclass{sigma}
\fi

 \usepackage{tikz-cd}

\numberwithin{equation}{section}

\newtheorem{Theorem}{Theorem}[section]
\newtheorem*{Theorem*}{Theorem}

\newtheorem{Lemma}[Theorem]{Lemma}
\newtheorem{Proposition}[Theorem]{Proposition}
 { \theoremstyle{definition}
\newtheorem{Definition}[Theorem]{Definition}

\newtheorem{Example}[Theorem]{Example}
\newtheorem{Construction}[Theorem]{Construction}
\newtheorem{Observation}[Theorem]{Observation}
\newtheorem{Remark}[Theorem]{Remark} }

\DeclareMathOperator*{\ran}{ran}

\DeclareMathOperator{\im}{\mathrm{im}}
\DeclareMathOperator{\dom}{\mathrm{dom}}
\DeclareMathOperator{\graph}{\mathrm{graph}}
\newcommand{\Lag}{\textsc{Lag}}
\newcommand{\LagRel}{\textsc{LagRel}}
\newcommand{\LagCor}{\textsc{LagCor}}
\newcommand{\pHilb}{\textsc{pHilb}}
\newcommand{\R}{\mathbb{R}}
\newcommand{\N}{\mathbb{N}}

\newcommand{\Z}{\mathbb{Z}}
\newcommand{\C}{\mathbb{C}}

\newcommand{\ind}{\operatorname{ind}}
\newcommand{\Cl}{\mathrm{Cl}}
\newcommand{\CCl}{\mathbb{C}\mathrm{l}}

\newcommand{\id}{\mathrm{id}}

\newcommand{\sBord}{\textsc{sBord}}

\begin{document}
\allowdisplaybreaks

\renewcommand{\thefootnote}{}

\newcommand{\arXivNumber}{2212.02956}

\renewcommand{\PaperNumber}{036}

\FirstPageHeading

\ShortArticleName{Categories of Lagrangian Correspondences in Super Hilbert Space}

\ArticleName{Categories of Lagrangian Correspondences in Super\\ Hilbert Spaces and Fermionic Functorial Field Theory\footnote{This paper is a~contribution to the Special Issue on Global Analysis on Manifolds in honor of Christian B\"ar for his 60th birthday. The~full collection is available at \href{https://www.emis.de/journals/SIGMA/Baer.html}{https://www.emis.de/journals/SIGMA/Baer.html}}}

\Author{Matthias LUDEWIG}

\AuthorNameForHeading{M.~Ludewig}

\Address{Fakult\"at f\"ur Mathematik, Universit\"at Regensburg, Germany}
\Email{\href{mailto:matthias.ludewig@mathematik.uni-regensburg.de}{matthias.ludewig@mathematik.uni-regensburg.de}}

\ArticleDates{Received May 19, 2023, in final form April 16, 2024; Published online April 24, 2024}

\Abstract{In this paper, we study Lagrangian correspondences between Hilbert spaces. A~main focus is the question when the composition of two Lagrangian correspondences is again Lagrangian. Our answer leads in particular to a well-defined composition law in a category of Lagrangian correspondences respecting given polarizations of the Hilbert spaces involved. As an application, we construct a functorial field theory on geometric spin manifolds with values in this category of Lagrangian correspondences, which can be viewed as a formal Wick rotation of the theory associated to a free fermionic particle in a curved spacetime.}


\Classification{19K56; 58J20; 81T45}

\begin{flushright}
\it Dedicated to the 60th birthday of Christian B\"ar
\end{flushright}

\renewcommand{\thefootnote}{\arabic{footnote}}
\setcounter{footnote}{0}

\section{Introduction}

In this article, we consider the category of Lagrangian correspondences relevant to quantization of fermionic fields.
While in the bosonic case, Lagrangian subspaces of \emph{symplectic} vector spaces are relevant, in the fermionic case, the symplectic form is replaced by a non-degenerate \emph{symmetric} bilinear form $B$ (\emph{Hermitian} in the complex case), and Lagrangians in this context are maximally isotropic subspaces for $B$.
Throughout, we take $V$ to be a Hilbert space and assume that~${B(x, y) = \langle \Gamma x, y\rangle}$ for an isometric involution $\Gamma$ of $V$.
Such an involution induces a~$\Z_2$-grading on $V$, hence this is the same set of data as a \emph{super} Hilbert space.

For functorial quantization, one would like to have a category whose objects are super Hilbert spaces as above, and whose morphisms are Lagrangian relations $L \subset \Pi V_0 \oplus V_1$, where $\Pi V_0$ is the super Hilbert space with the opposite grading (correspondingly, $B$ replaced by $-B$).
However, the composition of arbitrary Lagrangian relations may fail to be Lagrangian again, so this naive definition does not yield a category with a well-defined composition law.

A solution to this problem is to consider \emph{polarized} super Hilbert spaces instead, i.e., Hilbert spaces $V$ with a fixed equivalence class of Lagrangians (or, more generally, \emph{sub-Lagrangians}, see Definition~\ref{DefinitionLagrangian}) in $V$.
A Lagrangian correspondence $L \subset \Pi V_0 \oplus V_1$ is then required to be either the graph of an isometry that preserves the polarization, or equivalent to the Lagrangian $L_0^\perp \oplus L_1$, where $L_i$ defines the polarization of $V_i$ (such Lagrangians are called \emph{split}).
When restricting to this subclass of Lagrangians, on obtains a category $\LagRel$ of Lagrangian relations with a~well-defined composition law (see Section~\ref{SectionPolarizations}).

Our investigations are motivated by fermionic functorial field theory \cite{DaiFreed, FreedLectures, LudewigRoos, StolzTeichnerElliptic, StolzTeichnerSusy}.
Such a field theory assigns a Hilbert space $V_Y$ of fields to a (compact) spacelike slice $Y$ of spacetime, and to a region $X$ of spacetime bounding spacelike slices $Y_0$ and $Y_1$ a time evolution operator~$Q_X\colon V_{Y_0} \to V_{Y_1}$.
In our main example -- thoroughly described in this article --, $V_Y$~is the space of spinors on $Y$, which are polarized by the (equivalence class of the) Atiyah--Patodi--Singer sub-Lagrangian (see Definition~\ref{APSLagrangian}).
In this example, the corresponding time evolution operator~$Q_X$ is the solution operator to the Cauchy problem for the Dirac equation on $X$.
$Q_X$ is determined by requiring $Q \varphi_0 = \varphi_1$ if there exists a spinor $\Phi$ on $X$ with boundary values~${\Phi|_{Y_i} = \varphi_i}$ such that $D_X \Phi = 0$ \cite[Section~2]{BaerStrohmaier}.
In \emph{Lorentzian} signature, this is a unitary operator, and (identifying $Q_X$ with its graph), this construction can be viewed as a functor from a suitable Lorentzian spin bordism category to $\LagRel$, in which all morphisms occurring in the image are graph Lagrangians.

Things are very different in Riemannian signature: If $X$ is a \emph{Riemannian} bordism between~$Y_0$ and $Y_1$, there is no well-posed Cauchy problem.
The time evaluation operator can still be defined by $Q_X\varphi_0 = \varphi_1$ if $\varphi_i = \Phi|_{Y_i}$ for some spinor $\Phi$ on $X$ with $D_X \Phi = 0$.
However, in the Riemannian case, $Q_X$ is far from unitary:
It is unbounded, a reflection of the fact that the Cauchy problem is not solvable for each $\varphi_0$.
Nevertheless, its graph $L_X = \graph(Q_X)$ is still a~well-defined Lagrangian relation.
This gives a functorial field theory
\begin{equation}
\label{FieldTheoryIntro}
\begin{tikzcd}
 \mathcal{L}_0\colon\ \sBord_d \ar[r] & \LagRel
\end{tikzcd}
\end{equation}
defined on the Riemannian spin bordism category, which, in contrast to the case of Lorentzian signature, requires non-graphical Lagrangians in the target category $\LagRel$.
This theory was sketched in \cite[Remark~3.16]{LudewigRoos}.

We remark that in order for the \emph{Lorentzian} solution operator $Q_X$ to be well defined, we need the Lorentzian manifold $X$ to be globally hyperbolic, as this is the correct setting to solve wave equations \cite{BGP}.
However, by the famous theorem of Bernal--S\'anchez \cite{BernalSanchez}, such manifolds are diffeomorphic to $Y \times [0, 1]$, with $Y$ being the spacelike slice.
Hence there are no topologically interesting bordisms in the Lorentzian setting.
Again, this is in stark contrast to the \emph{Riemannian} setup, where many topologically distinct bordisms exist.
It is intriguing to think that this topological information is already contained in the Lorentzian field theory, but it only becomes visible after ``Wick rotation'', a point of view the author learned from Dan Freed.

From a physical point of view, the field theory described above is ``first-quantized'', i.e., describing a one-particle system.
Second quantization should then be given by postcomposing with the exponential functor
\begin{equation}
\label{SecondQuantizationDotted}
\begin{tikzcd}
 \LagRel \ar[r, dotted] & \textsc{sAlg}
\end{tikzcd}
\end{equation}
to the bicategory of superalgebras, bimodules and intertwiners,
which assigns to a Hilbert space~$V$ a corresponding Clifford algebra $\Cl(V)$ and to a Lagrangian correspondence ${L\! \subseteq\! \Pi V_0 \!\oplus\! V_1}$ the fermionic Fock space $\mathcal{F}_L = \Lambda L$, a $\Cl(V_1)$-$\Cl(V_0)$-bimodule.
This second-quantized theory was previously constructed in \cite{LudewigRoos}.
However, it was also observed there that the above second-quantization functor is ``anomalous'', in the sense that it is only projectively functorial, and that the field theory constructed in \cite{LudewigRoos} does \emph{not} factor through the functor \eqref{FieldTheoryIntro}.
Again, this ``chiral anomaly'' only appears when split Lagrangians are composed, and hence in our framework only occurs in Riemannian signature.

In this article, we solve this problem by introducing a larger category of \emph{Lagrangian correspondences}, where morphisms are spans
\begin{equation*}
\begin{tikzcd}[row sep = 0.2mm]
 & \mathcal{H} \ar[dl, "r_1"'] \ar[dr, "r_0"] &\\
 V_1 & & V_0,
\end{tikzcd}
\end{equation*}
such that the kernel of $(r_0, r_1)\colon \mathcal{H} \to \Pi V_0 \oplus V_1$ is finite-dimensional and the image is a Lagrangian (either split or graphical).
As there is an obvious notion of morphism between Lagrangian correspondences (isomorphisms of spans), this yields a bicategory, instead of an ordinary category.
The field theory \eqref{FieldTheoryIntro} can then be enhanced to a field theory
\begin{equation}
\label{FTEnhancementIntro}
 \mathcal{L} \colon\ \sBord_d \longrightarrow \LagCor
\end{equation}
taking values in $\LagCor$, which assigns to a bordism $X$ the vector space $\mathcal{H}_X$ of harmonic spinors on $X$, with $r_i$ the boundary restriction map $\Phi \mapsto \Phi|_{Y_i}$.
This field theory is described in Section~\ref{SectionFermionicFieldTheory}.
Moreover, with $\LagRel$ replaced by the larger bicategory $\LagCor$, the second quantization \eqref{SecondQuantizationDotted} now exists as an honest functor
\begin{equation*}
\begin{tikzcd}
 \textsf{Q}\colon\ \LagCor \ar[r] & \textsc{sAlg}.
\end{tikzcd}
\end{equation*}

The main contents of this paper are the following.
\begin{enumerate}\itemsep=0pt
\item[(1)]
In Section~\ref{SectionRelations}, we give a complete and self-contained description of the theory of Lagrangian relations $L \subset \Pi V_0 \oplus V_1$ and their composition, which so far was not available in the literature beyond the results of \cite[Section~2]{LudewigRoos}.
Particular features of our presentation are the description of Lagrangians and their composition in terms of unitaries (see Section~\ref{SectionLagrangianGraph} and Theorem~\ref{ThmCompositionUnitaryPicture}).
\item[(2)]
In Section~\ref{SectionCategory}, we give a complete description of a bicategory of Lagrangian correspondences, and, correspondingly, a solution to the anomaly problem of the second quantization functor~\eqref{SecondQuantizationDotted}.
This solves a problem posed in~\cite{LudewigRoos} and should be of general interest.
We also introduce versions of the correspondence category that take into account Clifford actions, which are naturally present in the field theory picture.
\item[(3)]
In Section~\ref{SectionFieldTheorySUP}, we give a construction of a field theory $\mathcal{L}$ from \eqref{FTEnhancementIntro}, whose main property is that it factors the field theory of \cite{LudewigRoos} as the composition
\begin{equation*}
 \begin{tikzcd}
 \sBord_d \ar[r, "\mathcal{L}"] & \LagCor \ar[r, "\textsf{Q}"] & \textsc{sAlg},
\end{tikzcd}
\end{equation*}
of a geometric/analytic and a purely algebraic part.
\end{enumerate}

\section{Lagrangians relations and their composition}\label{SectionRelations}

In this section, we study (sub-)Lagrangians and Lagrangian relations in abstract super Hilbert spaces.
The main result is Theorem~\ref{ThmCompositionUnitaryPicture}, which gives a sufficient criterion for the composition of two Lagrangian relations to be Lagrangian again.

\subsection{Lagrangians in super Hilbert spaces}

Let $V = V^+ \oplus V^-$ be a real or complex super Hilbert space with grading operator $\Gamma$, the isometric involution that takes the value $\pm 1$ on $V^\pm$.
Using this, we define the bilinear (respectively sequilinear) form
\begin{equation} \label{BilinearFormB}
B(x, y) := \langle \Gamma x, y\rangle = \langle x^+, y^+\rangle - \langle x^-, y^-\rangle, \qquad x, y \in V,
\end{equation}
where $x= x^+ + x^-$, $y = y^+ + y^-$ is the splitting with respect to the decomposition $V = V^+ \oplus V^-$.
A subspace $L \subset V$ is called \emph{$B$-isotropic} if $B(x, y) = 0$ for all $x, y \in L$.
We have the following simple lemma.

\begin{Lemma} \label{LemmaIntersectionZero}
For any $B$-isotropic subspace $L \subset V$, we have $L \cap V^+ = L \cap V^- = \{0\}$.
\end{Lemma}

\begin{proof}
Let $x \in L \cap V^\pm$.
Then since $L$ is $B$-isotropic, $B(x, x) = 0$.
On the other hand, since~${x \in V^\pm}$, we have $\Gamma x = \pm x$, so
\begin{equation*}
0 = B(x, x) = \langle \Gamma x, x\rangle = \pm \langle x, x\rangle = \pm\|x\|^2.\tag*{\qed}
\end{equation*}\renewcommand{\qed}{}
\end{proof}

\begin{Definition}[Lagrangians] \label{DefinitionLagrangian}
A \emph{sub-Lagrangian} is a $B$-isotropic subspace such that $L + \Gamma L$ has finite codimension in $V$.
$L$ is a \emph{Lagrangian} if this codimension is in fact zero, i.e., $L + \Gamma L = V$.
\end{Definition}

That $L$ is $B$-isotropic means equivalently that $\Gamma L \subseteq L^\perp$.
Such a $B$-isotropic subspace is a~sub-Lagrangian if $\Gamma L$ moreover has finite codimension in $L^\perp$ and a Lagrangian if $\Gamma L = L^\perp$.
In particular, sub-Lagrangians are always closed subspaces.
If $L$ is a Lagrangian in a super Hilbert space $V$, orthogonal projections $P_L$ and $P_{L^\perp}$ onto $L$ and its complement satisfy the relation
\begin{equation*}
 P_{L^\perp} = 1 - \Gamma P_L \Gamma.
\end{equation*}

\subsection{Lagrangians as graphs of operators}
\label{SectionLagrangianGraph}

For a closed operator $A \colon V \to W$ between Hilbert spaces, we denote by \[\graph(A) = \{(x, Ax) \mid x \in V\} \subset V \oplus W\] its graph, and
\begin{equation*}
\graph^\prime(A) = \graph(A) \cap \ker(A)^\perp,
\end{equation*}
where we freely identify $\ker(A) \subseteq V$ with the subspace $\ker(A) \oplus\{0\} \subseteq V \oplus W$.

\begin{Lemma} \label{LemmaLagrangianCorrespondence}
Let $V$ be a super Hilbert space.
For any closed $B$-isotropic subspace $L \subset V$, there exists a unique partial isometry $u\colon V^+ \to V^-$ such that $L = \graph^\prime(u)$.
Conversely, the graph of every such partial isometry is a closed $B$-isotropic subspace.
Moreover, a closed $B$-isotropic subspace $L = \graph(u)$ is a sub-Lagrangian if and only if $u$ is Fredholm and a Lagrangian if and only if $u$ is unitary.
\end{Lemma}

\begin{proof}
Let $L \subset V$ be a closed $B$-isotropic subspace.
Define the subspace $U \subset V^+$ by
\begin{equation*}
U = \big\{x^+ \in V^+ \mid \exists x^- \in V^- \colon x^+ + x^- \in L\big\}.
\end{equation*}
Define $\tilde{u}\colon U \to V^-$ by $\tilde{u}x^+ = x^-$ if $x^+ + x^- \in L$.
Observe that if $x^-_1, x^-_2 \in V^-$ are two elements such that $x_1= x^+ + x^-_1$ and $x_2 = x^+ + x^-_2$ are in $L$, then $x_1 - x_2 = x^-_1 - x^-_2 \in L \cap V^-$.
But this intersection is zero by Lemma~\ref{LemmaIntersectionZero}, hence $x^-_1 = x^-_2$ and $\tilde{u}$ is well defined.
It is linear because $L$ is a subspace.
For $x = x^+ + x^-$, $y = y^+ + y^-$ in $L$, using that $L$ is $B$-isotropic, we calculate
\begin{equation*}
 0 = B(x, y) = \big\langle x^+ - x^-, y^+ + y^-\big\rangle = \big\langle x^+, y^+\big\rangle - \big\langle x^-, y^-\big\rangle = \big\langle x^+, y^+\big\rangle - \big\langle \tilde{u}x^+, \tilde{u}y^+\big\rangle,
\end{equation*}
hence $\tilde{u}$ is an isometry.
This implies that $U$ is a closed subspace: if $x_n^+$, $n \in \N$ is a sequence of elements in $U$ converging to $x^+ \in V^+$, then since $u$ is an isometry, $ux_n^+$ converges to some~${y^- \in V^-}$.
Hence the sequence of elements $x_n^+ + u x_n^+$ in $L$ converges against $x^+ + y^- \in V$, which lies in $L$ since $L$ is closed.
Hence $x^+ \in U$ and $y^- = u x^+$.
Extending $\tilde{u}$ by zero on $U^\perp$ gives a partial isometry $u$ such that $\graph^\prime(u) = L$.

Conversely, let $u \colon V^+ \to V^-$ be a partial isometry.
Then for $x^+, y^+ \in \ker(u)^\perp \subset V^+$, we have
\begin{equation*}
 B\big(x^+ + u x^+, y^+ + u y^+\big) = \big\langle \Gamma\big(x^+ + u x^+\big), y^+ + u y^+ \big\rangle = \big\langle x^+, y^+\big\rangle - \big\langle ux^+, uy^+\big\rangle = 0,
\end{equation*}
as $u$ is isometric on $\ker(u)^\perp$.
Hence $\graph^\prime(u)$ is $B$-isotropic.
It is a closed subspace since it is the orthogonal complement of the closed subspace $\ker(u)$ inside the closed subspace $\graph(u)$.

For any $B$-isotropic subspace $L \subset V$, we have
\begin{equation} \label{LPlusGammaLperp}
(L \oplus \Gamma L)^\perp = (\graph^\prime(u) \oplus \graph^\prime(-u^*))^\perp = \ker(u) \oplus \ker(u^*).
\end{equation}
Since $u$ is a partial isometry, it has closed range.
Hence $u$ is Fredholm if and only if $\ker(u)$ and $\ran(u)^\perp = \ker(u^*)$ are finite-dimensional.
But by \eqref{LPlusGammaLperp} this is equivalent to $L$ being a sub-Lagrangian.
It also follows from \eqref{LPlusGammaLperp} that $L$ is a Lagrangian if and only if the partial isometry~$u$ is injective and surjective, in other words, unitary.
\end{proof}

\begin{Lemma} \label{LemmaOrthogonalComplement}
If $L = \graph^\prime(u)$ is a sub-Lagrangian in $V$, then $\Gamma L$ is a sub-Lagrangian and $\Gamma L = \graph^\prime(-u) = \graph^\prime(-u^*)$.
\end{Lemma}

\begin{proof}
Let $x^+ + ux^+ \in \graph(\pm u)$, $x^+ \in V^+$.
Then $\Gamma\big(x^+ ux^+\big) = x^+ \mp u x^+$.
Therefore, $\Gamma\graph(u) = \graph(-u)$.
Since $\ker(u) = \ker(-u)$, we also have $\graph^\prime(u) = \graph^\prime(-u)$.

To see that $\graph^\prime(-u) = \graph^\prime(-u^*)$, let $x^+ - u x^+ \in \graph^\prime(-u)$.
Then $x^+ \in \ker(u)^\perp = \ran(u^*)$, hence $x^+ = u^* y^-$ for some $y^- \in V^-$.
Here we can choose the preimage $y^-$ to lie in $\ker(u^*)^\perp$, so that $uu^* y^-= y^-$, as $u$ is a partial isometry.
So
\[x^+ - u x^+ = u^*y^- - u u^* y^- = u^*y^- - y^- \in \graph(-u^*).\]
This shows that $\graph^\prime(-u) \subseteq \graph^\prime(-u^*)$.
Swapping the roles of $u$ and $u^*$ gives the converse inclusion.
\end{proof}

\begin{Remark}
In the case that $L$ is a Lagrangian, to prove Lemma~\ref{LemmaOrthogonalComplement}, we can also use the general fact that for any closed, densely defined operator $A\colon V_0 \to V_1$ between Hilbert spaces, we have $\graph(A)^\perp = \graph(-A^*)$.

Hence if $L = \graph(u)$ is a Lagrangian, then $\Gamma L = \graph(u)^\perp = \graph(-u^*)$.
\end{Remark}

\begin{Lemma} \label{LemmaLagrangianContainingGraph}
If $L \subset V$ is a $B$-isotropic subspace and $\graph(u) \subset L$ for some operator $u\colon V^+ \to V^-$, then $u$ is an isometry and $L = \graph(u)$.
Moreover, we have $(L + \Gamma L)^\perp = \ker(u^*) \subset V^-$.
\end{Lemma}

\begin{proof}
As in the proof of Lemma~\ref{LemmaLagrangianCorrespondence}, one obtains that $u$ is an isometry.
For any $x = {x^+ + x^- \in L}$, we have $x^+ + u x^+ \in \graph(u) \subset L$, hence $x - (x^+ + u x^+) = x^- - u x^+ \in L \cap V^-$.
Hence $x^- = u x^+$, by Lemma~\ref{LemmaIntersectionZero}.
Therefore, $L = \graph(u)$.

To see that $(L + \Gamma L)^\perp = \ker(u^*)$, we use that $L^\perp = \graph(u)^\perp = \graph(-u^*)$.
Therefore, $(L + \Gamma L)^\perp$ consists of those $x \in \graph(-u^*)$ with $x \perp \Gamma L$.
Write $x = x^- - u^*x^-$ for $x^- \in V^-$ and $y \in \Gamma L$ as $y = y^+ + u y^+$.
Then $\langle x, y \rangle = 0$ if and only if
\begin{align*}
 0 &= \big\langle \Gamma\big(y^+ + u y^+\big), x^- - u^* x^-\big\rangle = \big\langle y^+ - u y^+, x^- - u^* x^- \big\rangle = - 2\big\langle y^+, u^* x^-\big\rangle.
\end{align*}
Therefore, $x \perp \Gamma L$ if and only if $x^- \in \ker(u^*)$.
\end{proof}

\begin{Remark}
One easily checks that if $L = \graph^\prime(u) \subset V$ is a sub-Lagrangian, then the orthogonal projection onto $L$ is given in terms of $u$ by
\begin{equation} \label{FormulaForProjection}
 P_L = \frac{1}{2} \begin{pmatrix} u^*u & u^* \\ u & uu^* \end{pmatrix}, \qquad \text{with reference to} \quad V = V^+ \oplus V^-.
\end{equation}
\end{Remark}

\subsection{Lagrangian relations}

For a super Hilbert space $V$ denote by $\Pi V$ the super Hilbert space with the opposite grading, in other words, the grading operator $\Gamma$ is replaced by its negative.
For two super Hilbert spaces~$V_0$ and $V_1$, we consider the sum $\Pi V_0 \oplus V_1$, which has
\begin{equation*}
(\Pi V_0 \oplus V_1)^+ = V_0^- \oplus V_1^+, \qquad (\Pi V_0 \oplus V_1)^- = V_0^+ \oplus V_1^-.
\end{equation*}

\begin{Definition}[Lagrangian relation]
A \emph{Lagrangian relation} between $V_0$ and $V_1$ is a Lagrangian~${L \subset \Pi V_0 \oplus V_1}$.
We say that a Lagrangian relation $L$ is in \emph{general position} if
\begin{equation*}
L \cap (V_0 \oplus \{0\}) = L \cap (\{0\} \oplus V_1) = \{0\}.
\end{equation*}
\end{Definition}

\begin{Remark}
Since $V_0 \oplus \{0\}$ and $\{0\} \oplus V_1$ are invariant under the grading operator $\Gamma$, and $\Gamma$ sends $L$ to its orthogonal complement, it follows that also $L^\perp$ has trivial intersection with these spaces.
The following result is therefore a consequence of the more general Theorem~1 of \cite{HalmosTwoSubspaces}, but it admits a shorter proof in the present context.
\end{Remark}

\begin{Lemma} \label{LemmaGraph1}
Let $L \subset \Pi V_0 \oplus V_1$ be a Lagrangian relation in general position.
Then $L$ is the graph of a closed, densely defined operator $T\colon V_0 \rightarrow V_1$ with zero kernel and dense range.
\end{Lemma}

\begin{proof}
Set
\begin{align*}
&\dom(T) = \{x_0 \in V_0 \mid \exists x_1 \in V_1\colon (x_0, x_1) \in L\}, \\
&\ran(T) = \{x_1\in V_1 \mid \exists x_0 \in V_0\colon (x_0, x_1) \in L\}.
\end{align*}
Then setting $Tx_0 = x_1$ if $(x_0, x_1) \in L$ defines a bijection between $\dom(T)$ and $\ran(T)$.
It is single-valued because if $(x_0, x_1), (x_0, x_1^\prime) \in L$, then also $(0, x_1 - x_1^\prime)\in L$, which implies that~${x_1 - x_1^\prime = 0}$, as $L$ is in general position.
Similarly, one sees that $T$ is injective.
By construction, ${\graph(T) = L}$, so $T$ is a closed operator.
Suppose now that $y \perp \dom(T)$ for some $y \in V_0$.
Then also $(y, 0) \perp (x_0, x_1)$ for all $(x_0, x_1) \in L$.
Since $L$ is a Lagrangian, this implies $(y, 0) \in \Gamma{L}$.
But because ${L}$, hence also $\Gamma{L}$, is in general position, this implies that $y=0$, hence $\dom(T)$ is dense.
The proof that $\ran(T)$ is dense is similar.
\end{proof}

In the lemma below, we use the following notation:
If $T\colon V_0 \to V_1$ is a linear operator between super Hilbert spaces with grading operators $\Gamma_0$, respectively $\Gamma_1$, we set
\begin{equation*}
 T^\Gamma := \Gamma_1 T \Gamma_0.
\end{equation*}
As follows from the proof of Lemma~\ref{LemmaGraph1}, the inverse $T^{-1}\colon V_1 \to V_0$ exists as a closed, densely defined operator.
The following statement relates this operator to $T$.

\begin{Lemma} \label{LemmaRelationTAdjointInverse}
Let $T\colon V_0 \to V_1$ be an invertible closed operator.
Then $L = \graph(T)$ is a~Lagrangian in $\Pi V_0 \oplus V_1$ if and only if
\begin{equation} \label{RelationTAdjointInverse}
 T^{-1} = (T^*)^\Gamma.
\end{equation}
\end{Lemma}

\begin{proof}
If $L = \graph(T)$ is a Lagrangian in $\Pi V_0 \oplus V_1$, we may calculate
\begin{align*}
 0 = B\bigl(x, Tx), (y, Ty)\bigr)
 &=\langle \Gamma(x, Tx), (y, Ty)\rangle = \langle (-\Gamma_0 x, \Gamma_1Tx), (y, Ty)\rangle \\
 &= - \langle \Gamma_0 x, y\rangle + \langle \Gamma_1 Tx, Ty \rangle = \langle (T^* \Gamma_1 T - \Gamma_0) x, y\rangle
\end{align*}
for every $x, y \in V_0$.
Hence $T^* \Gamma_1 T - \Gamma_0 = 0$, or $\Gamma_0 T^* \Gamma_1 = T^{-1}$.

Conversely, if $T$ is a closed, densely defined, invertible operator satisfying \eqref{RelationTAdjointInverse}, then
\begin{equation*}
\graph(T)^\perp = \graph(-T^*) = \graph\bigl(- \big(T^{-1}\big)^\Gamma\bigr) = (\Gamma_0, -\Gamma_1)\graph\bigl(T^{-1}\bigr).
\end{equation*}
But also $\graph(T) = \graph(T^{-1})$.
\end{proof}

Above, we have seen how to write a Lagrangian $L \subset \Pi V_0 \oplus V_1$ as the graph of an operator~${T\colon V_0 \to V_1}$.
On the other hand, by Lemma~\ref{LemmaLagrangianCorrespondence}, there exists a unitary
\begin{equation} \label{Matrixu}
 u = \begin{pmatrix} u_{00} & u_{01} \\ u_{10} & u_{11} \end{pmatrix}\colon\ (\Pi V_0 \oplus V_1)^+ = V_0^- \oplus V_1^+ \longrightarrow V_0^+ \oplus V_1^- = (\Pi V_0 \oplus V_1)^-.
\end{equation}
We will now relate these two points of view.

\begin{Lemma} \label{LemmaU01TrivialKernel}
Suppose that the Lagrangian $L = \graph(u)$ is in general position and write $u$ as in \eqref{Matrixu}.
Then both $u_{01}$ and $u_{10}$ have trivial kernel and dense range.
\end{Lemma}

\begin{proof}
Let $x \in V_1^+$ with $u_{01} x = 0$.
Then $(0, x+u_{11}x) \in L$, and since $L$ is by assumption in general position, this implies $x + u_{11}x = 0$, hence $x = 0$ \big(as $x \in V_1^+$ but $u_{11}x \in V_1^-$\big).
We obtain that $\ker u_{01} = \{0\}$.
Similarly, one shows that $\ker u_{10} = \{0\}$.

It remains to show that $\mathrm{ran}(u_{01})^{\perp} = \ker(u_{01}^*) = \{0\}$, and the same thing for $u_{10}$.
However, $L^{\perp} = \graph(-u^*)$ is a Lagrangian in $\Pi V$, which is in general position since $L$ is, so the previous argument applies to show that also $u_{10}^*$ and $u_{01}^*$ have trivial kernel.
\end{proof}

With respect to the splitting $V_i = V_i^{\pm}$, we may write
\begin{equation*}
 T = \begin{pmatrix} T_{++} & T_{+-} \\ T_{-+} & T_{--} \end{pmatrix}\colon\ V_0 = V_0^+ \oplus V_0^- \longrightarrow V_1^+ \oplus V_1^- = V_1.
\end{equation*}

\begin{Theorem} \label{TheoremRelationTu}
The following formulas express $T$ and $u$ in terms of each other:
\begin{equation*}
 T = \begin{pmatrix} u_{01}^{-1} & -u_{01}^{-1} u_{00} \\ u_{11} u_{01}^{-1} & u_{10} - u_{11} u_{01}^{-1}u_{00} \end{pmatrix}, \qquad
 u = \begin{pmatrix} - T_{++}^{-1} T_{+-} & T_{++}^{-1} \\ T_{--} - T_{-+}T_{++}^{-1} T_{+-} & T_{-+}T_{++}^{-1}\end{pmatrix}.
\end{equation*}
\end{Theorem}

\begin{proof}
Let $x \in V_0^-$, $y \in V_0^+$. Then the vectors
\begin{equation} \label{MembersOfL1}
 (x + u_{00}x, u_{10} x), \qquad (x, T_{+-}x + T_{--}x), \qquad (y, T_{++}y + T_{-+} y)
 \end{equation}
 are all contained in $L$.
 Setting $y = u_{00} x$ and adding the latter two expressions yields
 \begin{equation*}
 (x + u_{00}x, T_{+-}x + T_{--}x + T_{++}u_{00}x + T_{-+} u_{00}x) \in L.
 \end{equation*}
 Comparing with the first element of \eqref{MembersOfL1} and using that $L$ is in general position, we obtain
 \begin{equation} \label{EqTandU}
 T_{+-} + T_{++}u_{00} = 0 \qquad \text{and} \qquad T_{--} + T_{-+}u_{00} = u_{10}.
 \end{equation}
 Let now $x \in V_1^+$.
 Then we obtain $(u_{01}x, x + u_{11}x) \in L$.
 Comparing this with the third element of \eqref{MembersOfL1} (where we set $y = u_{01}x$), we obtain
 \begin{equation} \label{owef}
 T_{++}u_{01} = 1 \qquad \text{and} \qquad T_{-+} u_{01} = u_{11}.
 \end{equation}
 It follows from Lemma~\ref{LemmaU01TrivialKernel} that the inverse of $u_{01}$ exists as a (possibly unbounded) densely defined, closed operator, hence \eqref{owef} implies the result together with \eqref{EqTandU}.
\end{proof}

\begin{Example}
Suppose that $V_0 = V_1 = \C^2$ and that
\begin{equation*}
 u = \begin{pmatrix} \cos\alpha & - \sin\alpha \\ \sin\alpha & \cos\alpha \end{pmatrix}.
\end{equation*}
For the corresponding operator $T$, we then get
\begin{equation*}
 T = \begin{pmatrix} \frac{\cos\alpha}{\sin\alpha} & - \frac{1}{\sin\alpha} \\ \frac{1}{\sin\alpha} & \frac{\cos\alpha}{\sin\alpha} \end{pmatrix} = \begin{pmatrix} \cot\alpha & - \csc\alpha \\ \csc\alpha & \cot\alpha \end{pmatrix}.
\end{equation*}
\end{Example}

\begin{Lemma} \label{LemmaTunitary}
The operator $T$ is unitary if and only if $u$ is off-diagonal, i.e., $u_{11} = u_{00} = 0$.
In this case, $T$ is diagonal, i.e., $T_{+-} = T_{-+} = 0$, and $u_{10}$, $u_{01}$ are unitary.
\end{Lemma}

\begin{proof}
It is clear that if $u$ is off-diagonal, then $u_{10}$ and $u_{01}$ are unitary, and so is $T$, with $T_{+-} = T_{-+}=0$, by inspection of the formula for $T$ from Theorem~\ref{TheoremRelationTu}.

Conversely, if $T$ is unitary, then each of the four block operators $T_{+-}$, $T_{-+}$, $T_{++}$, $T_{--}$ is bounded, with operator norm less or equal to one.
In particular, we have $\|T_{+-}\| \leq 1$.
On the other hand, from the relation $u_{01} = T_{+-}^{-1}$, we obtain that also $\big\|T_{+-}^{-1}\big\| \leq 1$.
By definition of the operator norm, this means that $\|x\| \leq \|T_{+-}x\|$ for all $x \in V$, and the same is true for the adjoint~$T_{+-}^*$.
Hence $\|T_{+-}T_{+-}^*\| = \|T_{+-}^*\|^2 \leq 1$ and
\begin{equation*}
\langle x, T_{+-} T_{+-}^* x \rangle = \|T_{+-}^*x\|^2 \geq \|x\|^2.
\end{equation*}
In total, we obtain that $T_{+-} T_{+-}^* = 1$.
That $T$ is unitary implies the relation $T_{++} T_{++}^* + T_{+-} T_{+-}^* = 1$.
But this means that $T_{++} T_{++}^*= 0$, hence also $T_{++}=0$.
In a similar fashion, one shows that $T_{--} = 0$, hence $T$ is block diagonal.
By Theorem~\ref{TheoremRelationTu}, this implies that $u_{00} = u_{11} = 0$.
That $u_{10}$, $u_{01}$ are unitary then follows from the fact that $u$ is unitary.
\end{proof}

\subsection{Composition of Lagrangian relations}

\begin{Definition}
Let $V_0$, $V_1$, $V_2$ be three super Hilbert spaces
and let
\begin{equation*}
L_{01} \subset \Pi V_0 \oplus V_1 \qquad \text{and} \qquad L_{12} \subset \Pi V_1 \oplus V_2
\end{equation*}
 be two Lagrangian relations.
The \emph{composition} of $L_{01}$ and $L_{12}$ is simply their composition as relations,
\begin{equation*}
 L_{12} \circ L_{01} = \bigl\{ (x_0, x_2) \in \Pi V_0 \oplus V_2 \mid \exists x_1 \in V_1 \colon (x_0, x_1) \in L_{01}, (x_1, x_2) \in L_{02}\bigr\}.
\end{equation*}
\end{Definition}

\begin{Lemma} \label{LemmaCompositionIsotropic}
In the situation above, $L_{12} \circ L_{01}$ is $B$-isotropic in $\Pi V_0 \oplus V_2$.
\end{Lemma}

\begin{proof}
Let $(x_0, x_2), (y_0, y_2) \in \Pi V_0 \oplus V_2$ be two elements of $L_{12} \circ L_{01}$.
Then there exist $x_1, y_1 \in V_1$ such that $(x_0, x_1), (y_0, y_1) \in \Pi V_0 \oplus V_1$ and $(x_1, x_2), (y_1, y_2) \in \Pi V_1 \oplus V_2$.
We then have
\begin{align*}
B\bigl((x_0, x_2), (y_0, y_2)\bigr)
 &=\langle \Gamma_{02} (x_0, x_2), (y_0, y_2)\rangle = \bigl\langle (-\Gamma_0 x_0, \Gamma_2, x_2), (y_0, y_2) \bigr\rangle \\
 &= - \langle \Gamma_0 x_0, y_0\rangle + \langle \Gamma_2 x_2, y_2 \rangle + \langle \Gamma_1 x_1, y_1 \rangle - \langle \Gamma_1 x_1, y_1 \rangle\\
 &= \langle \Gamma_{01}(x_0, x_1), (y_0, y_1) \rangle + \langle \Gamma_{12} (x_1, x_2), (y_0, y_2) \rangle,
\end{align*}
where $\Gamma_i$ is the grading operator of $V_i$ and $\Gamma_{ij}$ denotes the grading operator of $\Pi V_i \oplus V_j$.
This is zero as $L_{01}$ and $L_{12}$ are $B$-isotropic.
\end{proof}

The following example illustrates that the composition need not be a Lagrangian, i.e., maximally isotropic.

\begin{Example} \label{ExampleNonComposable}
Consider the Hilbert space $V = \ell^2(\Z)$, with involution given by $\Gamma{e}_n = e_{-n}$, where~$e_n$,~$n \in \Z$, is the $n$-th standard basis vector.
For $\alpha \in \R$, consider the multiplication operator $T_\alpha$, given by $T_\alpha e_n = e^{\alpha n} e_n$, with $\dom(T_\alpha) = \big\{x \in \ell^2(\Z) \mid T_\alpha x \in \ell^2(\Z)\big\}$.
Then $T_\alpha$ is densely defined and closed.
One moreover easily checks that $T_\alpha$ is invertible, with $T^{-1}_\alpha =(T_\alpha^*)^\Gamma$, which implies by Lemma~\ref{LemmaRelationTAdjointInverse} that
\begin{equation*}
L_\alpha := \graph(T_\alpha)
\end{equation*}
is a Lagrangian in $\Pi V \oplus V$, i.e., a Lagrangian relation.
Clearly, the composition of two such Lagrangians $L_{\alpha_1}$ and $L_{\alpha_2}$ is the graph of the operator $T_{\alpha_1}T_{\alpha_2}$.
 When $\alpha_1$ and $\alpha_2$ have the same sign, then $T_{\alpha_1} T_{\alpha_2} = T_{\alpha_1+\alpha_2}$, so that $L_{\alpha_1} \circ L_{\alpha_2}$ is again Lagrangian.
 However, if $\alpha_1$ and $\alpha_2$ have opposite sign, we have $T_{\alpha_1}T_{\alpha_2} \subset T_{\alpha_1+\alpha_2}$, but the composition $T_{\alpha_1} T_{\alpha_2}$ is not closed (for example, if $\alpha_2 = - \alpha_1$, then $T_{\alpha_1} T_{\alpha_2} \subset \mathrm{id}$ but is not everywhere defined). Therefore, in the case that $\alpha_1$ and~$\alpha_2$ have opposite sign, the composition of $L_{\alpha_1}$ and $L_{\alpha_2}$ is not closed, hence not a Lagrangian.
 \end{Example}

 While in the example above, the composition of $L_{\alpha_1}$ and $L_{\alpha_2}$ may not be closed and therefore not maximal, at least its closure will always be maximal, hence Lagrangian.
 However more extreme phenomena are possible: After conjugating $T_{\alpha_2}$ by a suitable orthogonal transformation~$u$ of $\ell^2(\Z)$, the domain of $(u T_{\alpha_2}u^*) \circ T_{\alpha_1}$ can even be $\{0\}$, in other words, the composition of the corresponding Lagrangians is the zero subspace.

In the following, we will investigate under which conditions the composition $L_{12} \circ L_{01}$ is in fact a Lagrangian.
One result in this direction has been proved in \cite[Section~2.2, Theorem~2.11]{LudewigRoos}.
Here we prove a result using the representation of the Lagrangians as unitaries.
So let
\begin{align}
 u &= \begin{pmatrix} u_{00} & u_{01} \\ u_{10} & u_{11} \end{pmatrix} \colon\ V_0^- \oplus V_1^+ \longrightarrow V_0^+ \oplus V_1^-,\nonumber \\
 v &= \begin{pmatrix} v_{11} & v_{12} \\ v_{21} & v_{22} \end{pmatrix}\colon\ V_1^- \oplus V_2^+ \longrightarrow V_1^+ \oplus V_2^- \label{MatrixRepUV}
\end{align}
 be unitaries such that $L_{01} = \graph(u)$, $L_{12} = \graph(v)$, as in Lemma~\ref{LemmaLagrangianCorrespondence}.

\begin{Remark}
For later use, we record that the fact that $u$ and $v$ are unitary implies
\begin{alignat}{4}
 & 1 = u_{00}^* u_{00} + u_{10}^* u_{10}, \qquad && 1 = v_{11}^* v_{11} + v_{21}^* v_{21}, \qquad&&
 1 = u_{11}^* u_{11} + u_{01}^* u_{01}, &\nonumber \\
& 1 = v_{22}^* v_{22} + v_{12}^* v_{12}, \qquad&&
 0 = u_{00}^* u_{01} + u_{10}^* u_{11}, \qquad&& 0 = v_{11}^* v_{12} + v_{21}^* v_{22},&\nonumber \\
& 0 = u_{01}^* u_{11} + u_{11}^* u_{10}, \qquad&& 0 = v_{12}^* v_{11} + v_{22}^* v_{21}.&&&\label{uvUnitary}
\end{alignat}
\end{Remark}

Let $A\colon V \to W$ be a bounded operator between Hilbert spaces with closed range.
By the \emph{generalized inverse} of $A$, we mean the operator $X \colon W \to V$ with $\ker(X) = \ran(A)^\perp$ (and~${\ran(X) = \ker(A)^\perp}$) such that $XA = 1$ on $\ker(A)^\perp$ and $AX = 1$ on $\ran(A)$.
Such a generalized inverse can be constructed by splitting $V = \ker(A) \oplus \ker(A)^\perp$, $W = \ran(A) \oplus \ran(A)^\perp$ (using that $A$ has closed range) and observing that the restriction $A^\prime\colon \ker(A)^\perp \to \ran(A)$ is invertible; then $X$ is the inverse of $A^\prime$ on $\ran(A)$ and zero on $\ran(A)^\perp$.
Our result is now the following.

\begin{Theorem}
\label{ThmCompositionUnitaryPicture}
Suppose that the operators $1 - v_{11} u_{11}$ and $1- u_{11} v_{11}$ have closed range.
Then the composition $L_{12} \circ L_{01}$ is a Lagrangian
and hence the graph of a unitary transformation
\begin{equation*}
 w\colon\ (\Pi V_0 \oplus V_2)^+ = V_0^- \oplus V_2^+ \longrightarrow V_0^+ \oplus V_2^- = (\Pi V_0 \oplus V_2)^-.
\end{equation*}
This unitary is given by
\begin{equation} \label{FormulaForComposition}
w = \begin{pmatrix} u_{00} + u_{01}X v_{11} u_{10} & u_{01} X v_{12} \\
 v_{21}Y u_{10} & v_{22} + v_{21}Y u_{11}v_{12} \end{pmatrix},
\end{equation}
where $X$ and $Y$ are the generalized inverses of $1-v_{11} u_{11}$ and $1-u_{11}v_{11}$, respectively.
\end{Theorem}

\begin{Remark}
In particular, if both $1- v_{11} u_{11}$ and $1- u_{11} v_{11}$ are actually invertible, then
\begin{equation*}
w = \begin{pmatrix} u_{00} + u_{01}(1-v_{11}u_{11})^{-1} v_{11} u_{10} & u_{01} (1-v_{11}u_{11})^{-1} v_{12} \\
 v_{21}(1-u_{11}v_{11})^{-1} u_{10} & v_{22} + v_{21}(1-u_{11}v_{11})^{-1} u_{11}v_{12} \end{pmatrix}.
\end{equation*}
\end{Remark}

We need the following preliminary results.

\begin{Lemma} \label{LemmaKernelU11V11}
We have
\begin{align*}
\ker(1 - v_{11} u_{11}) &\subseteq \ker(u_{01}) \cap \ker(v_{21} u_{11}), \qquad \ker(1- u_{11}v_{11}) \subseteq \ker(u_{01} v_{11}) \cap \ker(v_{21}).
\end{align*}
\end{Lemma}

\begin{proof}
For $x \in V_1^+$, we have $(u_{01} x, x+u_{11}x) \in L_{01}$.
If now $v_{11} u_{11} x = x$,
then, since $u_{11} x \in V_1^-$, we have
\begin{equation*}
 (u_{11} x + v_{11} u_{11} x, v_{21} u_{11} x) = (x + u_{11} x, v_{21} u_{11} x) \in L_{12}.
\end{equation*}
Comparing the two elements just considered, we obtain that $(u_{01} x, v_{21} u_{11} x) \in L_{12} \circ L_{01}$.
However, we have $u_{01} x \in V_0^+$ and $v_{21}u_{11}x \in V_2^-$, hence $(u_{01} x, v_{21} u_{11} x) \in (\Pi V_0 \oplus V_2)^-$.
But since the composition $L_{12} \circ L_{01}$ is isotropic by Lemma~\ref{LemmaCompositionIsotropic}, this implies $(u_{01} x, v_{21} u_{11} x) = 0$, by Lemma~\ref{LemmaIntersectionZero}, so $x \in \ker(u_{01}) \cap \ker(v_{21} u_{11})$.

Similarly, if $x \in V_1^-$ with $x \in\ker(1 - u_{11} v_{11})$, we obtain $(x+v_{11}x, v_{21}x) \in L_{12}$ and
\begin{equation*}
 (u_{01} v_{11} x, v_{11} x + u_{11} v_{11} x) = (u_{01} v_{11} x, x + v_{11} x) \in L_{01}.
\end{equation*}
Hence \[(u_{01} v_{11} x, v_{21} x) \in (L_{12} \circ L_{01}) \cap (\Pi V_0 \oplus V_2)^-,\] which by Lemma~\ref{LemmaIntersectionZero} again implies $(u_{01} v_{11} x, v_{21} x) = 0$ and $x \in \ker(u_{01} v_{11}) \cap \ker(v_{21} u_{11})$.
\end{proof}

\begin{Remark}
If one of $L_{01}$ and $L_{12}$ is in general position, then either $u_{01}$ or $v_{12}$ has trivial kernel by Lemma~\ref{LemmaU01TrivialKernel}.
Hence in that case, it follows from Lemma~\ref{LemmaKernelU11V11} that both $1 - v_{11} u_{11}$ and~${1- u_{11}v_{11}}$ have trivial kernel.
\end{Remark}

\begin{Remark}
If $x \in \ker(u_{01}) \cap \ker(v_{21} u_{11})$, we get from \eqref{uvUnitary} the equations $x = u_{11}^* u_{11} x$ and~${u_{11} x = v_{11}^* v_{11} u_{11}x}$.
Put together, this gives $x = u_{11}^* v_{11}^* v_{11} u_{11} x$.
In general, this only implies that $x$ is an eigenvector for an eigenvalue $\lambda$ with $|\lambda| = 1$ of $v_{11} u_{11}$, but we do not necessarily have $\lambda = 1$.
It is easy to come up with explicit examples in two dimensions for $u$ and $v$ where this indeed occurs, so the inclusion of Lemma~\ref{LemmaKernelU11V11} is general proper.
\end{Remark}

\begin{Lemma} \label{LemmaComplementOfRange}
 We have
 \begin{align*}
 \ker(1 - v_{11} u_{11}) &\perp \ran(v_{12}) + \ran(v_{11} u_{10}),\qquad \ker(1 - u_{11} v_{11}) \perp \ran (u_{10}) + \ran (u_{11} v_{12}).
 \end{align*}
\end{Lemma}

\begin{proof}
Let $x \in \ker(1-v_{11} u_{11})$. Then by Lemma~\ref{LemmaKernelU11V11}, we have $v_{21} u_{11} x = 0$ and $u_{01} x= 0$. Hence for any $y \in V_2^+$, we have
\begin{equation*}
 \langle v_{12} y, x \rangle = \langle v_{12} y, v_{11} u_{11} x \rangle = \langle v_{11}^* v_{12} y, u_{11} x \rangle = - \langle v_{21}^* v_{22} u, u_{11} x\rangle = -\langle v_{22} u, v_{21} u_{11} x\rangle = 0.
\end{equation*}
Here we used the identities \eqref{uvUnitary} in the third step. Similarly, for any $y \in V_0^-$, we get
\begin{align*}
 \langle v_{11} u_{10} y, x\rangle = \langle v_{11} u_{10}, v_{11} u_{11} x\rangle &= \langle v_{11}^* v_{11} u_{10} y, u_{11} x \rangle = \langle (1- v_{21}^* v_{21}) u_{10} y, u_{11} x\rangle \\
 &= \langle u_{10} y, u_{11} x \rangle - \langle v_{21} u_{10} y, \underbrace{v_{21} u_{11} x}_{=0} \rangle\\
 &= \langle y, u_{10}^*u_{11} x \rangle = -\langle y, u_{00}^*u_{01} x \rangle = 0,
\end{align*}
where we used two of the identities \eqref{uvUnitary}.
This shows the first claim.
The proof of the second claim is similar.
\end{proof}

\begin{proof}[Proof of Theorem~\ref{ThmCompositionUnitaryPicture}]
Let $x^+ = \big(x_0^-, x_1^+\big) \in V_0^- \oplus V_1^+$ and $y^+ = \big(y_1^-, y_2^+\big) \in V_1^-\oplus V_2^+$.
Then $x^+ + ux^+ \in L_{01}$, $y^+ +vy^+ \in L_{12}$ can be expressed as
\begin{align}
& x^++ux^+= (x_0, x_1) = \big(\underbrace{u_{00} x_0^- + u_{01}x_1^+}_{ \in V_0^+} + \underbrace{x_0^-}_{\in V_0^-}, \underbrace{x_1^+}_{\in V_1^+} + \underbrace{u_{10} x_0^- + u_{11} x_1^+}_{\in V_1^-}\big),\nonumber\\
&y^++vy^+= (y_1, y_2) = \big(\underbrace{v_{11} y_1^- + v_{12}y_2^+}_{ \in V_1^+} + \underbrace{y_1^-}_{\in V_1^-}, \underbrace{y_2^+}_{\in V_2^+} + \underbrace{v_{21} y_1^- + v_{22} y_2^+}_{\in V_2^-}\big).\label{ElementsOfComposition}
\end{align}
By definition, the element $(x_0, y_2)$ is in the composition if $x_1 = y_1$.
Comparing the components of $x_1$ and $y_1$ in $V_1^+$ and $V_1^-$, the equation $x_1 = y_1$ gives the system of equations
\begin{equation*}
\left\{\begin{aligned}
x_1^+ &= v_{11} y_1^- + v_{12}y_2^+\\
y_1^- &= u_{10} x_0^- + u_{11} x_1^+
\end{aligned}\right\}
\quad \Longrightarrow \quad
\left\{\begin{aligned}
x_1^+ &= v_{11} (u_{10} x_0^- + u_{11} x_1^+) + v_{12}y_2^+\\
y_1^- &= u_{10} x_0^- + u_{11} ( v_{11} y_1^- + v_{12}y_2^+)
\end{aligned}\right\},
\end{equation*}
which can be rearranged to
\begin{align}
(1- v_{11}u_{11})x_1^+ &= v_{11} u_{10} x_0^- + v_{12}y_2^+, \label{Eqq1}\\
(1-u_{11}v_{11})y_1^- &= u_{10} x_0^- + u_{11}v_{12}y_2^+.\label{Eqq2}
\end{align}
By Lemma~\ref{LemmaComplementOfRange}, the right-hand side of \eqref{Eqq1} is orthogonal to $\ker(1- v_{11}u_{11})$, while the right-hand side of \eqref{Eqq2} is orthogonal to $\ker(1- u_{11}v_{11})$.
As $1- v_{11}u_{11}$ and $1-u_{11}v_{11}$ have closed ranges, their generalized inverses $X$ and $Y$ exist, which can by used to solve
 \begin{align}
x_1^+ &= Xv_{11} u_{10} x_0^- + Xv_{12}y_2^+,\qquad
y_1^- = Yu_{10} x_0^- + Yu_{11}v_{12}y_2^+.\label{Solutionx1y1}
\end{align}
We therefore get, substituting this into \eqref{ElementsOfComposition}, that given any choice of $x_0^- \in V_0^-$ and $y_2^+ \in V_2^+$, the element
\begin{align*}
&\big(\underbrace{u_{00} x_0^- + u_{01}X\big(v_{11} u_{10} x_0^- + v_{12}y_2^+\big)}_{ \in V_0^+} + \underbrace{x_0^-}_{\in V_0^-},\underbrace{y_2^+}_{\in V_2^+} + \underbrace{v_{21} Y\big(u_{10} x_0^- + u_{11}v_{12}y_2^+\big) + v_{22} y_2^+}_{\in V_2^-}\big)\\
&\qquad= \big(w_{00} x_0^- + w_{02} y_2^+ + x_0^-, y_2^+ + w_{20} x_0^- + w_{22} y_2^+\big)
\end{align*}
is the composition of $x^+ + u x^+$ and $y^+ + vy^+$ with $x^+ = \big(x_0^-, x_1^+\big)$, $y^+ = \big(y_1^-, y_2^+\big)$, where $x_1^+$ and $y_1^-$ are determined by $x_0^-$ and $y_2^+$ through \eqref{Solutionx1y1}.
This shows that the composition $L_{12} \circ L_{01}$ contains $\graph(w)$, where
\begin{equation*}
 w = \begin{pmatrix} w_{00} & w_{02} \\ w_{20} & w_{22} \end{pmatrix}\colon\ V_0^- \oplus V_2^+ \longrightarrow V_0^+ \oplus V_2^-,
\end{equation*}
with the entries $w_{00}$, $w_{02}$, $w_{20}$, $w_{22}$ as in \eqref{FormulaForComposition}.
By Lemma~\ref{LemmaLagrangianContainingGraph}, $w$ is an isometry and $L_{12} \circ L_{01} = \graph(w)$.
Moreover,
\[(L_{12} \circ L_{01} \oplus \Gamma(L_{12} \circ L_{01}))^\perp = \ker(w^*) \subset V_0^- \oplus V_2^+.\]

To see that $w$ is surjective, we can repeat the argument for the Lagrangian relations $L_{01}^{-1}$,~$L_{12}^{-1}$ in $\Pi V_1 \oplus V_0 $, respectively $\Pi V_2 \oplus V_1$ (here $L_{01}^{-1} = \{(x_1, x_0) \mid (x_0, x_1) \in L_{01}\}$).
These are then written as $L_{01}^{-1} = \graph(u^*)$, $L_{12}^{-1} = \graph(v^*)$.
Then by the same proof, their composition satisfies $L_{01}^{-1} \circ L_{12}^{-1} = \graph(\tilde{w})$ for some isometry $\tilde{w}\colon V_0^+ \oplus V_2^- \to V_0^- \oplus V_2^+$, and
\[(L_{12} \circ L_{01} \oplus \Gamma(L_{12} \circ L_{01}))^\perp = \ker(\tilde{w}^*) \subset V_0^+ \oplus V_2^-.\]
Together, we obtain that $(L_{12} \circ L_{01} \oplus \Gamma(L_{12} \circ L_{01}))^\perp = \{0\}$, $\ker(w^*) = \ker(\tilde{w}^*) = 0$ and $\tilde{w} = w^*$.
\end{proof}

\section{The category of Lagrangian correspondences}\label{SectionCategory}

In this section, we define bicategories of Lagrangian correspondences.

\subsection{Polarizations and split Lagrangians}\label{SectionPolarizations}

Given Hilbert spaces $V_0$ and $V_1$, we denote the space of Hilbert--Schmidt operators between $V_0$ and $V_1$ by $\mathcal{I}(V_0, V_1)$, or just $\mathcal{I}$ for brevity.
We remark that in most considerations below, the operator ideal of Hilbert--Schmidt operators can be replaced by any other symmetrically normed operator ideal.

\begin{Definition}[close subspaces]
Two closed subspaces $U$, $\tilde{U}$ of a Hilbert space $V$ are said to be {\em close}, or $U \sim \tilde{U}$, if the difference $P_{U} - P_{\tilde{U}} \in \mathcal{I}$, where $P_{U}$ and $P_{\tilde{U}}$ are the orthogonal projections onto $U$ and $\tilde{U}$, respectively.
\end{Definition}

Clearly, being close is an equivalence relation on the set of closed subspaces of $V$, which restricts to an equivalence relation on the set of sub-Lagrangians in $V$.
We write $[L]$ for the equivalence class of a sub-Lagrangian $L \subset V$ under closeness.

\begin{Definition}[polarized Hilbert spaces]
A {\em polarized super Hilbert space} is a (real or complex) super Hilbert space $V$ together with a distinguished equivalence class $[L]$ of sub-Lagrangians.
An {\em isomorphism} of polarized super Hilbert spaces $V_i = (V_i, [L_i])$, $i=0, 1$, is a grading-preserving unitary $T\colon V_0 \to V_1$ such that $TL_0 \in [L_1]$ for some (hence all) sub-Lagrangians $L_0 \subset V_0$ representing the polarization.
\end{Definition}

\begin{Example} \label{ExamplePolarizedHilbertSpace}
We have the following operations on polarized super Hilbert spaces:
\begin{enumerate}\itemsep=0pt
\item[(a)] The {\em opposite} of a super Hilbert space $V$ with polarization $[L]$ is the super Hilbert space $\Pi V$ with polarization $[\Gamma L]$.
\item[(b)] If $V_0 = (V_0, [L_0])$, $V_1 = (V_1, [L_1])$ are two polarized super Hilbert spaces, then a polarization on their direct sum is given by $[L_0 \oplus L_1]$.
Observe that the equivalence class~${[L_0 \oplus L_1]}$ contains all direct sum sub-Lagrangians $L_0^\prime \oplus L_1^\prime$ with $L_i^\prime \sim L_i$, but not all sub-Lagrangians ${L \in [L_0 \oplus L_1]}$ are of this form.
\end{enumerate}
\end{Example}

There are categories (in fact, groupoids) $\pHilb_{\R}$ and $\pHilb_{\C}$ with objects real, respectively complex, polarized super Hilbert spaces and isomorphisms.
These categories are symmetric monoidal with the direct sum operation from Example~\ref{ExamplePolarizedHilbertSpace}.

If $V_0$, $V_1$ are polarized super Hilbert spaces, combining Examples~\ref{ExamplePolarizedHilbertSpace}\,(a) and~(b) above, the polarized super Hilbert space $\Pi V_0 \oplus V_1$ is now defined, and be can consider Lagrangian relations which lie in the given polarization.

\begin{Definition}[types of Lagrangian relations]
Let $V_0$, $V_1$ be two polarized super Hilbert spaces and let $L \subset \Pi V_0 \oplus V_1$ be a Lagrangian relation.
\begin{enumerate}\itemsep=0pt
\item[(a)]
$L$ is called \emph{split} if $L$ is close to $L_0^\perp \oplus L_1$ for some (hence any) choice of sub-Lagrangians $L_i$ representing the polarization of $V_i$.
\item[(b)]
$L$ is called \emph{graphical} if $L = \graph(T)$ for an isomorphism $T\colon V_0 \to V_1$ of polarized super Hilbert space.
\end{enumerate}
\end{Definition}

Clearly, the composition of two graphical Lagrangians is graphical again, with
\[
\graph(T) \circ \graph(S) = \graph(TS).
\]
However, as seen in Example~\ref{ExampleNonComposable}, the composition of two general Lagrangian relations may fail to be Lagrangian again.
Propositions~\ref{PropCompositionOfSplit} and \ref{PropCompositionWithGraph} below show that this phenomenon cannot occur in the case that the Lagrangians are split and/or graphical.

\begin{Proposition}\label{PropCompositionOfSplit}
 Let $V_0$, $V_1$, $V_2$ be three polarized super Hilbert spaces and let
 \begin{equation*}
 L_{01} \subset \Pi V_0 \oplus V_1 \qquad \text{and} \qquad L_{12} \subset \Pi V_1 \oplus V_2
 \end{equation*}
 be two split Lagrangians.
 Then the composition $L_{12} \circ L_{01}$ is a Lagrangian in $\Pi V_0 \oplus V_2$, which is again split.
\end{Proposition}

For the proof, we need the following lemma.

\begin{Lemma} \label{LemmaClosenessForUnitaries}
Let $L_1, L_2 \subset V$ be sub-Lagrangians, which we write as $L_1 = \graph^\prime(u_1)$, $L_2 = \graph^\prime(u_2)$ for partial isometries $u_1, u_2\colon V^+ \to V^-$.
Then $L_1$ and $L_2$ are close if and only if~${u_1 - u_2 \in \mathcal{I}}$.
\end{Lemma}

\begin{proof}
By \eqref{FormulaForProjection}, $L_1$ and $L_2$ are close if and only if $u_1 - u_2$, $u_1^* - u_2^*$, $u_1^*u_1 - u_2^* u_2$ and $u_1 u_1^* - u_2 u_2^*$ are all in $\mathcal{I}$.
Hence it remains to show that $u_1 - u_2 \in \mathcal{I}$ implies that $u_1^*u_1 - u_2^* u_2$, $u_1 u_1^* - u_2 u_2^*$ are contained in $\mathcal{I}$.
We calculate
\begin{equation*}
u_1^* u_1 - u_2^* u_2 = (u_1^*u_1 - \id) - (u_2^* u_2 - \id).
\end{equation*}
This is of finite rank, since $L_1$ and $L_2$ are sub-Lagrangians, hence contained in $\mathcal{I}$.
The argument for $u_1 u_1^* - u_2 u_2^*$ is similar.
\end{proof}

\begin{proof}[Proof of Proposition~\ref{PropCompositionOfSplit}]
For $i = 0, 1, 2$, pick sub-Lagrangians $L_i = \graph^\prime(w_i)$ (with partial isometries $w_i\colon V_i^+ \to V_i^-$) defining the polarization of $V_i$ and write
$L_{01} = \graph(u)$ and ${L_{12} = \graph(v)}$, for unitaries $u$, $v$.
Moreover, write $u$ and $v$ as matrices as in \eqref{MatrixRepUV}.
It follows from Lemma~\ref{LemmaClosenessForUnitaries} that the split property of $L_{01}$ and $L_{12}$ implies that
\begin{equation} \label{ConditionTwo}
\begin{pmatrix} u_{00} & u_{01} \\ u_{10} & u_{11} \end{pmatrix} - \begin{pmatrix} -w_0^* & 0 \\ 0 & w_1 \end{pmatrix} \in \mathcal{I} \qquad \text{and}\qquad
\begin{pmatrix} v_{11} & v_{12} \\ v_{21} & v_{22} \end{pmatrix} - \begin{pmatrix} -w_1^* & 0 \\ 0 & w_2 \end{pmatrix} \in \mathcal{I}.
\end{equation}
In particular, all entries of these matrices are compact.
Therefore,
\begin{align*}
 1 - u_{11} v_{11} &= 2 - (u_{11} - w_1)v_{11} - w_1\bigl(v_{11} +w_1^*\bigr),\\
 1 - v_{11} u_{11} &= 2 - \bigl(v_{11} + w_1^*\bigr) u_{11} + w_1^*(u_{11} - w_1)
\end{align*}
are compact perturbations of a multiple of the identity, hence Fredholm operators.
In particular, they have closed ranges so that Theorem~\ref{ThmCompositionUnitaryPicture} yields that the composition $L_{12} \circ L_{01}$ is a~Lagrangian.
We have $L_{12} \circ L_{01} = \graph(w)$, with $w$ given by \eqref{FormulaForComposition}.
In order to verify that~${L_{12} \circ L_{01}}$ is close to $L_0^\perp \oplus L_2$, we need to check that
\begin{equation*}
\begin{pmatrix} u_{00} + u_{01}X v_{11} u_{10} & u_{01} X v_{12} \\
 v_{21}Y u_{10} & v_{22} + v_{21}Y u_{11}v_{12} \end{pmatrix}
 - \begin{pmatrix} -w_0^* & 0 \\ 0 & w_2 \end{pmatrix} \in \mathcal{I}.
\end{equation*}
But this is clear in view of the fact that $u_{00} + w_0^*$ and $v_{22} - w_2$ are in $\mathcal{I}$, as well as each of~$u_{01}X v_{11} u_{10}$,~$u_{01} X v_{12}$, $v_{21}Y u_{10}$ and $v_{21}Y u_{11}v_{12}$, as the off-diagonal terms of $v$ and $u$ are all compact.
\end{proof}

\begin{Proposition}
\label{PropCompositionWithGraph}
Let $V_0$, $V_1$, $W_0$ and $W_1$ be polarized super Hilbert spaces and let $L \subset \Pi V_0 \oplus V_1$ be a split Lagrangian.
Let moreover $S\colon V_0 \to W_0$ and $T\colon V_1 \to W_1$ be isomorphisms of polarized super Hilbert spaces.
Then the composition
\begin{equation*}
\graph(T) \circ L \circ \graph(S^*) \subset \Pi W_0 \oplus W_1
\end{equation*}
is a split Lagrangian.
\end{Proposition}

\begin{proof}
Assume first that $S$ is the identity.
Then $L \circ \graph(S^*) = L$ is a split Lagrangian, so we only have to show that $\graph(T) \circ L$ is a split Lagrangian.
Write $\graph(T) = \graph(u)$ and~${L = \graph(v)}$ for unitaries $u$ and $v$ as in \eqref{MatrixRepUV}.
Since $T$ is unitary, Lemma~\ref{LemmaTunitary} yields that~${u_{11} = u_{00} = 0}$.
In particular, $1-u_{11}v_{11}$ and $1-v_{11}u_{11}$ are both identities (in particular have closed ranges), so that $\graph(T) \circ L$ is a Lagrangian by Theorem~\ref{ThmCompositionUnitaryPicture}.

We need to show that $\graph(T) \circ L$ is split.
To this end, pick sub-Lagrangians $L_i = \graph^\prime(w_i)$ defining the polarization of $V_i$.
That $T L_0 \sim L_1$ means, by definition, that $T^*P_{L_1}T - P_{L_0} \in \mathcal{I}$.
By Theorem~\ref{TheoremRelationTu} (which expresses $T$ in terms of $u$) and the formula \eqref{FormulaForProjection} for the projections~$P_{L_i}$, this is equivalent to the requirement that
\begin{align*}
 &\begin{pmatrix} u_{01} & 0 \\ 0 & u_{10}^* \end{pmatrix}
 \begin{pmatrix}w_1^* w_1 & w_1^* \\ w_1 & w_1 w_1^* \end{pmatrix}
 \begin{pmatrix} u_{01}^* & 0 \\ 0 & u_{10} \end{pmatrix}
 -
 \begin{pmatrix}w_0^*w_0 & w_0^* \\ w_0 & w_0 w_0^* \end{pmatrix} \\
 &\qquad=
 \begin{pmatrix} u_{01} w_1^* w_1 u_{01}^* - w_0^* w_0 & u_{01}w_1^*u_{10} - w_0^* \\ u_{10}^*w_1 u_{01}^* - w_0& u_{01} w_1 w_1^* u_{10} - w_0 w_0^* \end{pmatrix} \in \mathcal{I}.
\end{align*}
Since $L_1$ and $L_0$ are sub-Lagrangians, the projections $w_0^* w_0$, $w_0 w_0^*$, $w_1^*w_1$ and $w_1 w_1^*$ differ from the identity by a finite rank operator, hence in particular are contained in $1 + \mathcal{I}$.
Therefore, ${u_{01} w_1^* w_1 u_{01}^* - w_0^* w_0}$ and $u_{01} w_1 w_1^* u_{10} - w_0 w_0^*$ are always in $\mathcal{I}$.
We arrive at the condition
\begin{gather} \label{ConditionOne}
u_{01}w_1^* u_{10} - w_0 \in \mathcal{I}.
\end{gather}
Using the formula from Theorem~\ref{ThmCompositionUnitaryPicture} for the representation of the composition and the observations in the beginning, we have to check that
\begin{equation*}
 \begin{pmatrix} u_{01} v_{11} u_{10} & u_{01} v_{12} \\
 v_{21} u_{10} & v_{22} + v_{21} u_{11}v_{12} \end{pmatrix}
 -
 \begin{pmatrix} -w_0^* & 0 \\
 0& w_2 \end{pmatrix} \in \mathcal{I}.
\end{equation*}
But this follows from \eqref{ConditionTwo} and \eqref{ConditionOne}.

For non-trivial $S$, we first use that by the above, $L^\prime := \graph(T) \circ L$ is a split Lagrangian in~${\Pi V_0 \oplus W_1}$.
The proof that $L^\prime \circ \graph(S^*)$ is a split Lagrangian in $\Pi W_0 \oplus W_1$ is then similar to the above.
\end{proof}

\begin{Definition}
\label{DefinitionCatLagRel}
The \emph{category of Lagrangian relations} $\LagRel$ has objects polarized Hilbert spaces and morphisms from $V_0$ to $V_1$ are Lagrangian relations $L \subset \Pi V_0 \oplus V_1$, which are either split or graphical.
\end{Definition}

By Propositions~\ref{PropCompositionOfSplit} and \ref{PropCompositionWithGraph}, composition of relations endows the category $\LagRel$ with a~well-defined composition law.

\subsection{Clifford actions}

Denote by $\Cl_{d}$ the real Clifford algebra of degree $d$, generated by $d$ anti-commuting elements $e_1, \dots, e_d$ with $e_j^2 = -1$.
Write $\CCl_d = \Cl_d \otimes \C$ for the complex Clifford algebra.
$\Cl_d$ and $\CCl_d$ have a natural $\Z_2$-grading (defined by declaring the generators $e_j$ to be odd), which turns them into superalgebras.
We write $\Cl_d^+$, respectively $\CCl_d^+$ for the even subalgebras, which are canonically isomorphic to $\Cl_{d-1}$, respectively $\CCl_{d-1}$, via the isomorphism given on the generators by
\begin{equation} \label{IsomorphismCliffordDimensionShift}
\Cl_{d-1} \ni e_j \mapsto e_d \cdot e_j \in \Cl_d^+, \qquad j=1, \dots, d-1.
\end{equation}
$\Cl_d$ is also real $*$-algebra, with $*$-operation defined by
\begin{equation*}
 (x_1 \cdots x_k)^* = x_k \cdots x_1.
\end{equation*}
Complex antilinear extension of this operation turns $\CCl_d$ into a complex $*$-algebra.

\begin{Definition}[degree $d$ Hilbert space]
A \emph{degree $d$ Hilbert space} is a (real or complex) super Hilbert space $H$ that is at the same time a graded right module for the Clifford algebra $\Cl_d$ such that the Clifford generators $e_j$ act as odd, skew-adjoint operators.
An isomorphism of degree $d$ Hilbert spaces is an isomorphism of super Hilbert spaces that intertwines the Clifford actions.
\end{Definition}

By a \emph{sub-Lagrangian} in a degree $d$ Hilbert space $V$, we mean an ordinary sub-Lagrangian~${L \!\subset\! L}$ that is invariant under the $\Cl_d$-action.
A \emph{polarization} on a degree $d$ Hilbert space is an equivalence class of $\Cl_d$-invariant sub-Lagrangians, where the equivalence relation does not take the Clifford action into account.
We write $\pHilb^d_\R$ and $\pHilb^d_\C$ for the categories of real, respectively complex polarized degree $d$ Hilbert spaces.

\begin{Example}
As $\Cl_1 \cong \C$, a real degree $1$ Hilbert space is the same as a complex Hilbert space.
The grading operator $\Gamma$ anti-commutes with the Clifford generator, which means that it is a real structure for the complex structure.
A Lagrangian $L$ is then a complex subspace such that its complex conjugate $\Gamma L$ equals its orthogonal complement.
This is the setting considered, e.g., in~\cite[Section~2.2]{StolzTeichnerElliptic}, \cite{LudewigRoos}.
\end{Example}

Following Lawson and Michaelsohn \cite{LawsonMichelsohn}, we denote by $\hat{\mathfrak{M}}_d$ (respectively $\hat{\mathfrak{M}}_d^\C$) the Grothendieck group of isomorphism classes of finitely generated, graded (right) $\Cl_d$-modules ($\CCl_d$-modules), with the group structure given by direct sum.
Elements of \smash{$\hat{\mathfrak{M}}_d^{(\C)}$} can be written has formal differences $[E] - [F]$, where $E$ and $F$ are finitely generated, graded $\Cl_d$-modules ($\CCl_d$-modules).
Similarly, denote by \smash{$\mathfrak{M}_d^{(\C)}$} the Grothendieck group of isomorphism classes of finitely generated ungraded $\Cl_d$-modules ($\CCl_d$-modules).
For any graded $\Cl_d$-module $E = E^+ \oplus E^-$, $E^+$ is an ungraded $\Cl_d^+ \cong \Cl_{d-1}$-module (for $d \geq 1$); this provides an equivalence between the categories of graded $\Cl_d$-modules and ungraded $\Cl_{d-1}$-modules \cite[Proposition~5.20]{LawsonMichelsohn}.
In particular this gives a canonical isomorphisms
\begin{equation*} 
 \hat{\mathfrak{M}}_d \cong \mathfrak{M}_{d-1}\qquad \text{and} \qquad \hat{\mathfrak{M}}^\C_d \cong \mathfrak{M}^\C_{d-1},
\end{equation*}
see \cite[Proposition~5.1]{AtiyahBottShapiro}.
Restriction along the inclusion $\Cl_d \to \Cl_{d+1}$ induces group homo\-morphisms \smash{$i^*\colon \mathfrak{M}_{d+1}^{(\C)} \to \mathfrak{M}_d^{(\C)}$}, which appear in the Atiyah--Bott--Shapiro isomorphisms
\begin{align*}
 \mathfrak{M}_{d-1} / i^* \mathfrak{M}_d \cong \hat{\mathfrak{M}}_d / i^* \hat{\mathfrak{M}}_{d+1} &\cong KO^{-d}(*), \qquad
 \mathfrak{M}^\C_{d-1} / i^* \mathfrak{M}^\C_d \cong \hat{\mathfrak{M}}^\C_d / i^* \hat{\mathfrak{M}}^\C_{d+1} \cong KU^{-d}(*),
 \end{align*}
see \cite{AtiyahBottShapiro} or \cite[Theorem~I.9.27]{LawsonMichelsohn}.

\begin{Definition}[index of sub-Lagrangian]
Let $L$ be a $\Cl_d$-invariant sub-Lagrangian in a real (complex) degree $d$ Hilbert space.
Its \emph{index} is the class
\begin{equation*}
 \ind(L) := \big[(L \oplus \Gamma L)^\perp\big] \in \hat{\mathfrak{M}}_d^{(\C)} / i^* \hat{\mathfrak{M}}_{d+1}^{(\C)}.
\end{equation*}
\end{Definition}

Observe here that the grading operator, as well as the action of $\Cl_d$ preserve $L \oplus \Gamma L$, hence also its orthogonal complement, so that $(L \oplus \Gamma L)^\perp$ is a (finite-dimensional) graded $\Cl_d$-module.

\begin{Example}
In particular, Lagrangians always have index zero and any subspace of a Lagrangian with finite codimension also has index zero.
To see the latter, let $L = L^\prime \oplus K \subseteq V$ be a~Lagrangian, with $K$ finite-dimensional.
Then $\ind(L^\prime)$ is represented by the graded $\Cl_d$-module~${K \oplus \Gamma K}$.
The $\Cl_d$-action extends to an action of $\Cl_{d+1}$, by letting $e_{d+1}$ act as
\begin{equation*}
 e_{d+1} = \begin{pmatrix} 0 & -\Gamma \\ \Gamma & 0 \end{pmatrix}.
\end{equation*}
Hence $K \oplus \Gamma K$ represents the zero class in $\hat{\mathfrak{M}}_d^{(\C)} / i^* \hat{\mathfrak{M}}_{d+1}^{(\C)}$.
\end{Example}

\begin{Lemma}
Two equivalent sub-Lagrangians $L_1$, $L_2$ have the same index.
\end{Lemma}

\begin{proof}
 Let $L_1 = \graph^\prime(u_1)$, $L_2 = \graph^\prime(u_2)$ for partial isometries $u_1, u_2 \colon V^+ \to V^-$.
 As~$L_1$ and $L_2$ are sub-Lagrangians, ${u}_1$ and ${u}_2$ are Fredholm (Lemma~\ref{LemmaLagrangianCorrespondence}).
 Because $\graph(u)^\perp = \graph(-u^*)$, we have
\begin{align*}
V &= \graph (u) \oplus \graph(-u^*) = \graph^\prime(u) \oplus \ker(u) \oplus \graph^\prime(-u^*) \oplus \ker(-u^*)\\
&= L \oplus \Gamma L \oplus \ker(u) \oplus \ker(u^*).
\end{align*}
 Define the odd, self-adjoint, $\Cl_d$-linear operators
 \begin{equation*}
 \tilde{u}_1 = \begin{pmatrix} 0 & u_1^* \\ u_1 & 0 \end{pmatrix}, \qquad \tilde{u}_2 = \begin{pmatrix} 0 & u_2^* \\ u_2 & 0 \end{pmatrix}.
 \end{equation*}
 Then clearly,
 \begin{equation*}
 \ind(L_i) = [\ker(\tilde{u}_i)] \in \hat{\mathfrak{M}}^{(\C)}_d / i^*\hat{\mathfrak{M}}^{(\C)}_{d+1}.
 \end{equation*}
 These are the graded indices of the odd, self-adjoint, $\Cl_d$-linear operators $\tilde{u}_1$ and $\tilde{u}_2$, as defined in \cite[Section~III.10]{LawsonMichelsohn}.
 By Lemma~\ref{LemmaClosenessForUnitaries}, we have $\tilde{u}_1 - \tilde{u}_2 \in \mathcal{I}$.
 In particular, the difference $\tilde{u}_1 - \tilde{u}_2$ is compact.
 Therefore, the operator $(1-t) \tilde{u}_1 + t \tilde{u}_2 = \tilde{u}_1 - t(\tilde{u}_1 - \tilde{u}_2)$ is Fredholm for each $t \in [0, 1]$, hence $\tilde{u}_1$ and $\tilde{u}_2$ lie in the same connected component of the space of odd, self-adjoint, $\Cl_d$-linear Fredholm operators.
 Since the $\Cl_d$-linear index is locally constant (see \cite[Proposition~III.10.6]{LawsonMichelsohn}), we have $\ind(u_1) = \ind(u_2)$.
\end{proof}

\begin{Example}
In the case that $d=0$, the group \smash{$\mathfrak{M}_0^{(\C)} / i^* \mathfrak{M}_1^{(\C)}$} is canonically isomorphic to~$\Z$ and under this isomorphism, the integer $\ind(L)$ for a sub-Lagrangian $L = \graph^\prime(u)$ is just given by the ordinary index $\ind(u) = \dim\ker(u) - \dim\ker(u^*)$ of the Fredholm operator~$u$.
\end{Example}

By the above lemma, the following is well defined.

\begin{Definition}[index of a polarized Hilbert space] \label{DefinitionIndexHilbertSpace}
The \emph{index} $\ind(V)$ of a degree $d$ Hilbert space $V$ with polarization $[L]$ is the index of the sub-Lagrangian $L^\prime$, for any $L^\prime \in [L]$.
\end{Definition}

To begin with, it is straight-forward to prove that the composition $L_{12} \circ L_{01}$ is a $\Cl_d$-invariant subspace of $\Pi V_0 \oplus V_2$ (this has independent of the split property).

\subsection{Lagrangian correspondences}

The following is a slight variation on the definition that was suggested in \cite[Definition~2.2.9]{StolzTeichnerElliptic}.

\begin{Definition}[generalized Lagrangians]
\label{DefinitionGeneralizedLagrangians}
Let $V$ be a degree $d$ Hilbert space.
A \emph{generalized Lagrangian} in $V$ consists of an (ungraded) $\Cl_{d}$-module $\mathcal{H}$ together with a $\Cl_d$-equivariant linear map
\begin{equation*}
r\colon\ \mathcal{H} \longrightarrow V,
\end{equation*}
such that $\im(r) \subset V$ is a Lagrangian in $V$ and such that $\ker(r)$ is finite-dimensional.
The \emph{defect} of the generalized Lagrangian $\mathcal{H} \stackrel{r}{\to} V$ is the dimension of the kernel of $r$.
If $V_0$ and~$V_1$ are degree~$d$ Hilbert spaces, a \emph{Lagrangian correspondence} between $V_0$ and $V_1$ is generalized Lagrangian~${\mathcal{H} \stackrel{r}{\to} \Pi V_0 \oplus V_1}$, which we view as a span
\begin{equation*}
\begin{tikzcd}[row sep = 0.2mm]
 & \mathcal{H} \ar[dl, "r_1"'] \ar[dr, "r_0"] &\\
 V_1 & & V_0,
\end{tikzcd}
\end{equation*}
 where $r_i \colon \mathcal{H} \to V_i$ is the map obtained by postcomposing $r\colon \mathcal{H} \to \Pi V_0 \oplus V_1$ with the projection onto the $i$-th factor.
 We say that $\mathcal{H}$ is \emph{split} if its image is a split Lagrangian.
\end{Definition}

\begin{Remark}
\label{RemarkInducedTopology}
 We do not assume any topology given on $\mathcal{H}$.
 However, since $\ker(r)$ is finite-dimensional, there is exactly one vector space topology on $\mathcal{H}$ that is Hausdorff, makes $r$ continuous and restricts to the standard topology on $\ker(r)$.
\end{Remark}

\begin{Remark}
In \cite[Definition~2.2.9]{StolzTeichnerElliptic}, it is only required that \emph{the closure} of the image of $r$ is a Lagrangian in $V$.
However, this causes problems with the composition operation discussed below.
In any case, if $\tilde{\mathcal{H}} \stackrel{r}{\to} V$ only satisfies the weaker condition of \cite{StolzTeichnerElliptic}, we may just take the completion with respect to the vector space topology from Remark~\ref{RemarkInducedTopology} and extend $\tilde{r}$ by continuity.
This will give a generalized Lagrangian in the sense of Definition~\ref{DefinitionGeneralizedLagrangians}.
\end{Remark}

\begin{Example}
 Of course, any Lagrangian $L \subset V$ is also a generalized Lagrangian, with $r\colon L \to V$ being the inclusion map.
 More generally, for any finite-dimensional $\mathbb{K}$-vector space~$K$, $\mathcal{H} := L \oplus K$ is a generalized Lagrangian with $r\colon \mathcal{H} \to V$ being the inclusion map of the first factor.
\end{Example}

 For any degree $d$ Hilbert space $V$, there is a category (in fact a groupoid) $\Lag^d(V)$ whose objects are generalized Lagrangians in $V$ and whose morphisms are vector space isomorphisms $\mathcal{H} \to \tilde{H}$ making the diagram
 \begin{equation*}
\begin{tikzcd}[row sep = 0.2mm]
 \mathcal{H} \ar[dr, "r"] \ar[dd, "\cong"']&\\
 & V_0 \\
\tilde{\mathcal{H}} \ar[ur, "\tilde{r}"'] &
\end{tikzcd}
\end{equation*}
commute.
Note that there is at most one morphism between $\mathcal{H}$ and $\mathcal{H}^\prime$ in case that $r$ is injective.

\begin{Proposition}\label{PropFiberProduct}
 Suppose we are given three polarized degree $d$ Hilbert spaces $V_0$, $V_1$, $V_2$ and generalized Lagrangians
 \begin{equation*}
\begin{tikzcd}[row sep = 0.2mm]
 & \mathcal{H}_{12} \ar[dl, "r_2"'] \ar[dr, "r_1"] & & \mathcal{H}_{01} \ar[dl, "r_1"'] \ar[dr, "r_0"] \\
 V_2 & & V_1 & & V_0
 \end{tikzcd}
\end{equation*}
whose images are split.
 Then their fiber product
 \begin{equation*}
\begin{tikzcd}[row sep = 0.2mm]
&& \mathcal{H}_{12} \times_{V_1} \mathcal{H}_{01} \ar[dl, dashed] \ar[dr, dashed] \\
 & \mathcal{H}_{12} \ar[dl, "r_2"'] \ar[dr, "r_1"] & & \mathcal{H}_{01} \ar[dl, "r_1"'] \ar[dr, "r_0"] \\
 V_2 & & V_1 & & V_0
 \end{tikzcd}
\end{equation*}
over $V_1$ is a generalized Lagrangian in $\Pi V_0 \oplus V_2$ whose image is split.
\end{Proposition}

We need the following lemma.

\begin{Lemma}
\label{LemmaOnK}
Given three polarized degree $d$ Hilbert spaces $V_0$, $V_1$, $V_2$ and split Lagrangians $L_{01} \subset \Pi V_0 \oplus V_1$, $L_{12} \subset \Pi V_1 \oplus V_2$.
Then
\begin{equation}
\label{DefinitionK}
K := \big\{ x \in V \mid (0, x) \in L_{01}, ~ (x, 0) \in L_{12} \big\}
\end{equation}
is finite-dimensional.
\end{Lemma}

\begin{proof}
Given $x \in K$, split $x = x^+ + x^-$ according to $V_1 = V_1^+ \oplus V_1^-$.
Then writing $L_{01} = \graph(u)$ and $L_{12} = \graph(v)$ for unitaries $u$, $v$ as in \eqref{MatrixRepUV}, we see that $x^- = u_{11} x^+$ and $x^+ = v_{11} x^-$.
Hence $x$ is a solution to
\begin{equation*}
(1 - v_{11}u_{11})x^+ = 0, \qquad (1 - u_{11}v_{11})x^- = 0.
\end{equation*}
We obtain that $K \subset \ker(1 - v_{11}u_{11}) \oplus \ker(1 - u_{11}v_{11})$. (In fact, $K$ is the graph of $u_{11}$ inside this direct sum).
But in the proof of Proposition~\ref{PropCompositionOfSplit}, we saw that both these operators are Fredholm operators, hence have a finite-dimensional kernel.
\end{proof}

\begin{proof}[Proof of Proposition~\ref{PropFiberProduct}]
 Let
 \begin{equation*}
 \mathcal{H} =
 \big\{ (\Phi_{01}, \Phi_{12}) \mid r_1(\Phi_{01}) = r_1(\Phi_{12})\big\}
 \end{equation*}
 be the fiber product.
 It comes with a map $r\colon\mathcal{H} \to \Pi V_0 \oplus V_2$ given by
 \begin{equation*}
 r(\Phi_{01}, \Phi_{12}) = \big(r_0\Phi_{01}), r_2(\Phi_{12})\big),
 \end{equation*}
 which we claim turns $\mathcal{H}$ into a generalized Lagrangian.
 First of all, it easy to see that $\im(r) = \im(r_{12}) \circ \im(r_{01})$, hence is a split Lagrangian by Proposition~\ref{PropCompositionOfSplit}.

 It remains to show that $\ker(r)$ is finite-dimensional.
 Let $(\Phi_{01}, \Phi_{12}) \in \ker(r)$ and write ${x = r_1(\Phi_{01}) = r_1(\Phi_{12})}$.
 Then $x \in K$, where $K$ is given by \eqref{DefinitionK}.
 But if $(\Phi_{01}^\prime, \Phi_{12}^\prime) \in \ker(r)$ is another element such that $x$ equals $r_1(\Phi_{01}^\prime) = r_1(\Phi_{12}^\prime)$, we have $\Phi_{01}^\prime - \Phi_{01} \in \ker(r_{01})$ and~${\Phi_{12}^\prime - \Phi_{12} \in \ker(r_{12})}$.
 We therefore have a short exact sequence
 \begin{equation*}
 \begin{tikzcd}
 0 \ar[r] & \ker(r_{01}) \oplus \ker(r_{12}) \ar[r] & \ker(r) \ar[r, "r_1"] & K \ar[r] & 0.
 \end{tikzcd}
 \end{equation*}
 The left space is finite-dimensional since $\mathcal{H}_{01}$ and $\mathcal{H}_{12}$ are generalized Lagrangians, while $K$ is finite-dimensional by Lemma~\ref{LemmaOnK}.
 Hence $\ker(r)$ must by finite-dimensional as well.
\end{proof}

\begin{Remark}
As seen in the proof above, the interesting feature of the fiber product of two generalized split Lagrangians $\mathcal{H}_{01}$ and $\mathcal{H}_{01}$ is that the defect of the resulting Lagrangian~${\mathcal{H}_{12} \times_{V_1} \mathcal{H}_{01}}$ may be larger than the sum of the defects of $\mathcal{H}_{12}$ and $\mathcal{H}_{01}$.
\end{Remark}

For polarized degree $d$ Hilbert spaces $V_0$, $V_1$, denote by $\LagCor^d(V_0, V_1)$ be the full subcategory of $\Lag^d(\Pi V_0 \oplus V_1)$ containing all generalized split Lagrangians and all graphical Lagrangians.
For three polarized degree $d$ Hilbert spaces $V_0$, $V_1$, $V_2$, there is a composition functor
\begin{equation*}
 \LagCor^d(V_1, V_2) \times \LagCor^d(V_0, V_1) \longrightarrow \LagCor^d(V_0, V_2),
\end{equation*}
which takes generalized Lagrangians to their fiber product.
This is well defined by Propositions~\ref{PropFiberProduct} and \ref{PropCompositionWithGraph}.
On morphisms, the image of two morphisms of spans is provided by the universal property of the fiber product.
This yields a bicategory $\LagCor^d$ with objects polarized degree $d$ Hilbert spaces and morphisms categories $\LagCor^d(V_0, V_1)$.
We write $\LagCor^d_\R$ or $\LagCor^d_\C$ for this bicategory if we want to emphasize the field we are working over.

\begin{Definition}
 We call the bicategory $\LagCor^d$ described above the \emph{bicategory of Lagrangian correspondences}.
\end{Definition}

\begin{Remark}
Allowing graphical Lagrangians as morphisms in addition to split Lagrangians is necessary in order for the category $\LagCor^d$ to have identity morphisms.
\end{Remark}

$\LagCor^d$ admits a forgetful functor to the bicategory of spans in the category of vector spaces, and the associator and unitor natural transformations are carried over from there.
In particular, the identity morphisms are the identity correspondences.
There is also a functor
\begin{equation*}
 \pHilb^d \longrightarrow \LagCor^d
\end{equation*}
from the category of polarized degree $d$ Hilbert spaces that is the identity on objects and assigns to an isomorphism of polarized degree $d$ Hilbert spaces its graph.
Here the left is an ordinary category, considered as a bicategory with only identity 2-morphisms, and the functoriality is strict.

\subsection{The second quantization functor}
\label{SectionSecondQuantization}

In this section, we describe second quantization functors
\begin{equation*}
 \textsf{Q}\colon\ \LagCor^d \longrightarrow \textsc{sAlg}
\end{equation*}
from the bicategory of Lagrangian correspondences defined above to the bicategory of superalgebras, bimodules and intertwiners (for any $d \in \N_0$), and both over $\R$ and $\C$.
This mainly relies on the results of \cite[Section~2]{LudewigRoos}.

We start with the real case.
Here the functor is defined by
\[
 \textsf{Q}(V) = \Cl(V, B),\]
the real algebraic Clifford algebra on $V$ with respect to the symmetric bilinear form $B$ defined in \eqref{BilinearFormB}.
Explicitly, this is the quotient of the tensor algebra on $V$ by the ideal generated by the Clifford relations
\begin{equation*}
 x \cdot y + y \cdot x = B(x, y), \qquad x, y \in V.
\end{equation*}
For a Lagrangian correspondence $\mathcal{H} \stackrel{r}{\to} \Pi V_0 \oplus V_1$, we set
\begin{equation*}
 \textsf{Q}(\mathcal{H}) = \Lambda \im(r) \otimes \Lambda^{\mathrm{top}} \ker(r),
\end{equation*}
the algebraic exterior algebra on the image of $r$, tensored with the determinant line of the (finite-dimensional) space $\ker(r)$.
This is a $\Cl(V_1)$-$\Cl(V_0)$-bimodule, which follows from the more general fact that whenever $L \subset V$ is a Lagrangian, then the exterior algebra $\Lambda L$ as a~left~$\Cl(V)$-module structure determined on generators by requiring that the action of $v \in L$ is given by the wedge product and $\Gamma v$ acts by contracting using $B$.

The main non-trivial statement is now that whenever
\[\mathcal{H}_{01} \to \Pi V_0 \oplus V_1\qquad \text{and}\qquad \mathcal{H}_{12} \to \Pi V_1 \oplus V_2\]
 are two Lagrangian correspondences, then there is a canonical isomorphism
\begin{equation}
\label{CoherenceIsomorphism}
 \textsf{Q}(\mathcal{H}_{12} \times_{V_1} \mathcal{H}_{01}) \longrightarrow \textsf{Q}(\mathcal{H}_{12}) \otimes_{\mathsf{Q}(V_1)} \textsf{Q}(\mathcal{H}_{01}),
\end{equation}
of $\textsf{Q}(V_2)$-$\textsf{Q}(V_0)$-bimodules, functorial in the $\mathcal{H}_{ij}$ and coherent for each composable triple of Lagrangian correspondences.
This is the (pseudo-)functoriality of the bicategorical functor $\textsf{Q}$, and is proven in \cite[Theorems~2.15 and~3.22]{LudewigRoos}.

In the complex case, we have the problem that $B$ is not symmetric but Hermitian, so that the corresponding Clifford algebra does not exist.
We therefore set instead
\begin{equation*}
 \mathsf{Q}(V) = \Cl\big(V \oplus \overline{V}, \tilde{B}\big),
\end{equation*}
where $\tilde{B}$ is the (now complex \emph{bilinear}) form
\begin{equation*}
 \tilde{B}((x_1, \overline{x}_2), (y_1, \overline{y}_2)) = B(x_2, y_1) + B(y_2, x_1).
\end{equation*}
In fact, this algebra is canonically isomorphic to the algebra of canonical anticommutation relations on $V$ for the Hermitian form $B$.
For a Lagrangian correspondence $\mathcal{H} \stackrel{r}{\to} \Pi V_0 \oplus V_1$, we set
\begin{equation*}
 \textsf{Q}(\mathcal{H}) = \Lambda \im(r) \otimes \overline{\Lambda \im(r)^\perp} \otimes \Lambda^{\mathrm{top}} \ker(r)\otimes \overline{\Lambda^{\mathrm{top}} \ker(r)},
\end{equation*}
which is again a $\textsf{Q}(V_2)$-$\textsf{Q}(V_0)$-bimodule.
This uses the fact that
\begin{equation*}
L = \im(r) \oplus \overline{\im(r)^\perp} \subset V \oplus \overline{V}
\end{equation*}
is a maximally isotropic subspace for the bilinear form $\tilde{B}$, i.e., satisfies $\tilde{\Gamma} L = L^\perp$, where $\tilde{\Gamma}$ is the \emph{conjugate} linear involution of $V \oplus \overline{V}$ given by \[\tilde{\Gamma}(x_1, \overline{x}_2) = \big(\Gamma x_1, \overline{\Gamma x_2}\big).\]
Just as in the real case, there are coherence isomorphisms as in \eqref{CoherenceIsomorphism} making this assignment (pseudo-)functorial.

\subsection{A symmetric monoidal correspondence category}
\label{SectionSymmetricMonoidalVersion}

Direct sum does \emph{not} provide a symmetric monoidal structure on the bicategory $\LagCor^d$ defined above, as in the case of infinite-dimensional spaces, the direct sum of a split Lagrangian and a~graph Lagrangian is neither split nor graph but a mixture of both, and hence not a morphism in this category.

One might try to fix this issue as follows:
Instead of insisting that the morphisms $\mathcal{H} \stackrel{r}{\to} \Pi V_0 \oplus V_1$ be either split or graphical, we might require the existence of direct sum decompositions $V_i = V_i^\prime \oplus V_i^{\prime\prime}$ such that $\mathcal{H}$ splits as a direct sum of a generalized split Lagrangian $\mathcal{H}^\prime$ in $\Pi V_0^\prime \oplus V_1^\prime$ and a graphical Lagrangian $\mathcal{H}^{\prime\prime} = \graph(T)$ for $T\colon V_0^{\prime\prime} \to V_1^{\prime\prime}$.
But when one allows arbitrary such direct sum decompositions, then the resulting ``bicategory'' does not possess well-defined composition functors.

We will now describe a solution to this problem, where morphisms are equipped with an extra structure and the morphisms are more restricted, in order to achieve well-defined composition functors.
To begin with, we consider the following category.

\begin{Definition}
Denote by $\pHilb^d_\otimes$ the category given as follows.
\begin{enumerate}\itemsep=0pt
\item[(1)]
Its objects are polarized degree $d$ Hilbert spaces $V$ that come with a fixed direct sum decomposition $V = \bigoplus_{i \in I} V_i$ as additional data, where $I$ is a finite index set.
In other words, each of the $V_i$ is a polarized degree $d$ Hilbert space and $V$ carries the direct sum polarization.
\item[(2)]
If $V = \bigoplus_{i \in I} V_i$ and $W = \bigoplus_{j \in J} W_j$ are two such objects, then a morphism $T\colon V \to W$ is an isomorphism of polarized degree $d$ Hilbert spaces such that each summand $V_i$ is mapped isomorphically to some summand $W_j$.
In other words, there exists a bijection $\kappa\colon I \to J$ and~$T$ is the direct sum of isomorphisms $T_i \colon V_i \to W_{\kappa(i)}$ of polarized degree $d$ Hilbert spaces.
\end{enumerate}
$\pHilb^d_\otimes$ has an obvious symmetric monoidal structure given by direct sum, where, if $I$ and $J$ are the index sets of the direct sum decompositions of $V$ and $W$, then $V \otimes W$ has a direct sum decomposition indexed by $I \sqcup J$.
\end{Definition}

The desired symmetric monoidal bicategory of generalized Lagrangians is obtained from $\pHilb^d_\otimes$ by adding certain morphisms and 2-morphisms.
Let \smash{$V = \bigoplus_{i \in I} V_i$} and \smash{$W = \bigoplus_{j \in J} W_j$} two objects of $\pHilb^d_\otimes$ and denote by
\begin{equation}
\label{LagOtimesMorphisms}
\LagCor^d_{\otimes}(V, W) \subseteq \Lag^d(\Pi V_0 \oplus V_1)
\end{equation}
the full subcategory consisting of those generalized Lagrangians $\mathcal{H}$ for which there exist subsets~${I^\prime \subseteq I}$ and $J^\prime \subseteq J$ such that $\mathcal{H}$ splits as a direct sum $\mathcal{H} = \mathcal{H}^\prime \oplus \mathcal{H}^{\prime\prime}$, where $\mathcal{H}^\prime$ is a generalized split Lagrangian in $\Pi V_I \oplus W_J$, with
 \begin{equation*}
 V_I := \bigoplus_{i \in I} V_i, \qquad W_J := \bigoplus_{j \in J} W_j,
 \end{equation*}
 and $\mathcal{H}^{\prime\prime} = \graph(T)$ for an isomorphism
 \[T\colon \ V_{I \setminus I^\prime} \to W_{J \setminus J^\prime}\]
 in $\pHilb^d_\otimes$.
 In other words, split Lagrangians are allowed to mix summands, but graphical Lagrangians are not.
 Using Propositions~\ref{PropFiberProduct} and \ref{PropCompositionWithGraph}, it is straight forward to show that the fiber product of generalized Lagrangians provides well-defined composition functors
 \begin{equation*}
 \LagCor^d_{\otimes}(V_1, V_2) \times\LagCor^d_{\otimes}(V_0, V_1) \longrightarrow \LagCor^d_{\otimes}(V_0, V_2).
 \end{equation*}

\begin{Definition}
 The bicategory $\LagCor^d_{\otimes}$ has objects the objects of $\pHilb^d$ and the categories \eqref{LagOtimesMorphisms} as morphism categories.
 We refer to it as the \emph{symmetric monoidal bicategory of Lagrangian correspondences}.
\end{Definition}

The direct sum of polarized degree $d$ Hilbert spaces and generalized Lagrangians equips $\LagCor^d_\otimes$ with a symmetric monoidal structure which extends that of $\pHilb_\otimes^d$ along the inclusion functor
\begin{equation}
\label{InclusionPHilbTimes}
 \pHilb^d_\otimes \longrightarrow \LagCor^d_\otimes
\end{equation}
that is the identity on objects and sends an isomorphism $T$ to its graph.
Here the left-hand side is viewed as a symmetric monoidal bicategory with only identity 2-morphisms.
Providing the full structure and checking all coherence conditions for this symmetric monoidal structure is long and tedious and will be omitted here; compare \cite{Schommer-Pries}.

\section{A fermionic field theory}
\label{SectionFieldTheorySUP}

In this section, we construct a functorial field theory with target the symmetric monoidal bicategory of Lagrangian correspondences, which is defined on Riemannian spin or spin$^c$ manifolds.

\subsection{Polarized Hilbert spaces from spin manifolds}
Let $X$ be a Riemannian manifold of dimension $d$.
Let $\Cl(X)$ be the corresponding bundle of Clifford algebras built from the tangent bundle.
Let $\CCl(X) = \Cl(X) \otimes \C$ be its complexification.
We use the following definition.

\begin{Definition}[spin structure] \label{DefinitionSpinStructure}
 A \emph{spin structure} on $X$ is a bundle $\Sigma_X$ of real graded $\Cl(X)$-$\Cl_d$-bimodules that is fiberwise a Morita equivalence (equivalently, irreducible).
 A \emph{spin$^c$ structure} on $X$ is a bundle $\Sigma_X$ of complex $\Z_2$-graded $\CCl(X)$-$\CCl_d$-bimodules that is fiberwise a Morita equivalence.
 We require $\Sigma_X$ to carry a metric and compatible connection such that the Clifford multiplication by vectors is skew-adjoint and satisfies the product rule
 \begin{equation*}
 \nabla^\Sigma_\xi(\eta \cdot \psi \cdot v) = \nabla_\xi \eta \cdot \psi \cdot v + \xi \cdot \big(\nabla_\eta^\Sigma \psi\big) \cdot u
 \end{equation*}
 for vector fields $\xi$, $\eta$ on $M$, sections $\psi$ of $\Sigma_X$ and vectors $v \in \R^d$.
 Here on the right-hand side, $\nabla$~denotes the Levi-Civita connection of~$TX$.
 A \emph{spin$^{(c)}$ manifold} is a~Riemannian manifold with a~spin$^{(c)}$-structure.
\end{Definition}

Here the condition that $\Sigma_X$ is a graded bimodule means that $\Sigma_X$ is a super vector bundle with a bimodule structure such that Clifford multiplication by vectors is odd.
We require the metric to make the even and odd parts orthogonal, and the connection is assumed to make the grading operator $\Gamma_X$ parallel.

\begin{Definition}[opposite]
The \emph{opposite} $X^\vee$ of a spin$^{(c)}$ manifold $X$ has the same underlying Riemannian metric and spinor bundle $\Sigma_X$, but the grading operator is replaced by its negative, $\Gamma_{X^\vee} = - \Gamma_X$, and the Clifford multiplication (temporarily denoted by $\bullet$) acquires a sign as well compared to the one of $X$,
\begin{equation*}
 \xi \bullet \psi = - \xi \cdot \psi, \qquad \psi \bullet v = - \psi \cdot v, \qquad \xi \in TX, \quad v \in \R_d, \quad\psi \in \Sigma_X.
\end{equation*}
\end{Definition}

\begin{Definition}[spin isometry]
Let $X$ and $Y$ be spin$^{(c)}$ manifolds.
A \emph{spin isometry} between~$X$ and $Y$ consists of an isometry of Riemannian manifolds $f\colon X \to Y$ together with bundle isomorphism $F\colon \Sigma_X \to \Sigma_Y$ covering $f$ that preserves the grading, metric and connection and intertwines the Clifford actions (more precisely, the left action is intertwined along the bundle isomorphism $\Cl(X) \to \Cl(Y)$ induced by $f$).
We will usually omit the map $F$ in notation, although it is always part of the data.
\end{Definition}

If $f\colon X \to \tilde{X}$ is a spin isometry, we get an induced spin isometry between the opposite manifolds,
\[ 
f^\vee\colon X^\vee \to \tilde{X}^\vee,
\]
which has the same underlying maps as $f$.

\begin{Construction}[opposite]
The \emph{opposite} $X^\vee$ of a spin$^{(c)}$ manifold $X$ has the same underlying Riemannian metric and spinor bundle $\Sigma_X$, but the grading operator is replaced by its negative, $\Gamma_{X^\vee} = - \Gamma_X$, and the Clifford multiplication (temporarily denoted by $\bullet$) acquires a sign as well compared to the one of $X$,
\begin{equation*}
 \xi \bullet \psi = - \xi \cdot \psi, \qquad \psi \bullet v = - \psi \cdot v, \qquad \xi \in TX, \quad v \in \R_d,\quad \psi \in \Sigma_X.
\end{equation*}
\end{Construction}

\begin{Construction}[boundary spin structure]
Let $X$ be a $d$-dimensional spin$^{(c)}$ manifold with boundary.
Then $X$ induces a spin$^{(c)}$-structure on $\partial X$ as follows.
The spinor bundle is
\begin{equation*}
 \Sigma_{\partial X} = \Sigma_X^+|_{\partial X},
\end{equation*}
with the grading operator of given by
\begin{equation} \label{BoundaryGradingOperator}
\Gamma\psi := \nu \cdot \psi \cdot e_d,
\end{equation}
where $\nu$ is the \emph{outward pointing} normal vector.
$\Sigma_{\partial X}$ is a bundle of graded $\Cl(\partial X)$-$\Cl_{d-1}$-modules with the Clifford multiplication (for the moment denoted by $\bullet$) defined by
\begin{equation*}
 \xi \bullet \psi = \nu \cdot \xi \cdot \psi, \qquad \psi \bullet v = \psi \cdot e_d \cdot v, \qquad \xi \in T\partial X, \quad v \in \R_{d-1},\quad \psi \in \Sigma_{\partial X}.
\end{equation*}
Observe, in particular, that by the definition \eqref{BoundaryGradingOperator}, Clifford multiplication is skew-adjoint, both with vectors of $T\partial X$ from the left and with vectors in $\R^{d-1}$ from the right.
Any spin isometry~${h\colon X \to X^\prime}$ between spin$^{(c)}$ manifolds with boundary induces a spin isometry $h|_{\partial X}\colon \partial X \to \partial X^\prime$ between the boundaries.
\end{Construction}

A spin$^{(c)}$ manifold $X$ has an associated \emph{Dirac operator} $D_X$, which is the first order differential operator acting on sections of $\Sigma_X$ defined by the formula
\begin{equation*}
 D_X \psi = \sum_{j=1}^d b_j \cdot \nabla^\Sigma_{b_j} \psi,
\end{equation*}
where $b_1, \dots, b_d$ is a local orthonormal basis of the tangent bundle $TX$.
Since Clifford multiplication is parallel, the Dirac operator is $\Cl_d$-linear, meaning that for any smooth section $\psi$ of~$\Sigma_X$, we have
\begin{equation*}
 D_X(\psi \cdot v) = (D_X \psi) \cdot v, \qquad v \in \R^d.
\end{equation*}
The Dirac operator is odd, i.e., it exchanges the chirality of sections of $\Sigma_X$.
Hence $D_X$ is the direct sum of its graded components
\begin{equation*}
 D^{\pm}_X \colon\ C^\infty\big(X, \Sigma^\pm_X\big) \longrightarrow C^\infty\big(X, \Sigma^\mp_X\big).
\end{equation*}

Let $Y$ be a $(d-1)$-dimensional spin$^{(c)}$ manifold.
We denote by
\begin{equation*}
V_{Y} = L^2(Y, \Sigma_Y)
\end{equation*}
the real super Hilbert space of square-integrable sections of $\Sigma_Y$, which is a graded right $\Cl_{d-1}$-module as well, hence a real degree $d-1$ Hilbert space.
In the case that $Y$ is closed, we obtain a polarization of $V_Y$ using the Dirac operator $D_Y$ of $Y$, as follows.
Since $Y$ is closed, elliptic regularity implies that $D_Y$ has discrete real spectrum with each eigenvalue of finite multiplicity.

\begin{Definition}
\label{DefinitionAPSLagrangian}
The \emph{Atiyah--Patodi--Singer sub-Lagrangian} is
\begin{equation} \label{APSLagrangian}
 L_Y = \overline{\bigoplus_{\lambda > 0} \mathrm{Eig}(D_Y, \lambda)}\subset V_Y,
\end{equation}
the Hilbert space direct sum of all eigenspaces to positive eigenvalues of $D_Y$.
\end{Definition}

As $D_Y$ is $\Cl_{d-1}$-linear, $L_Y$ is invariant under the $\Cl_{d-1}$-action.
As $D_Y$ is an odd operator, it follows that if $\psi$ is an eigenspinor of $D_Y$ to some eigenvalue $\lambda \in \R$, then $\Gamma \psi$ is an eigenspinor to the eigenvalue $-\lambda$.
We obtain that $\Gamma L_Y$ is the direct sum of all eigenspaces to negative eigenvalues, hence $L_Y \perp \Gamma L_Y$ and $(L_Y + \Gamma L_Y)^\perp = \ker(D_Y)$.
Since this is a finite-dimensional subspace, $L_Y$ is a sub-Lagrangian and hence defines a polarization of $V_Y$.
In total, $V_Y$ is a~polarized Hilbert space of degree $d$.

We summarize the above discussion as follows.

\begin{Construction} \label{ConstructionVX}
For any closed spin$^{(c)}$ manifold $Y$ of dimension $d-1$, we obtain a polarized degree $d-1$ Hilbert space $V_Y = (V_Y, [L_Y])$, where $V_Y$ is the space of square-integrable sections of the spinor bundle and $L_Y$ is the Atiyah--Patodi--Singer sub-Lagrangian \eqref{APSLagrangian}.
It comes with the additional data of a direct sum decomposition
\begin{equation}
\label{ObservationMonoidality}
 V_Y = V_{Y_1} \oplus \dots \oplus V_{Y_n},
\end{equation}
where $Y = Y_1 \sqcup \dots \sqcup Y_n$ is the decomposition into connected components.
$V_Y$ is a real Hilbert space in the spin case and a complex Hilbert space in the spin$^c$ case.

If $f\colon Y \to \tilde{Y}$ is a spin isometry, we obtain a spin isometry
\begin{equation}
\label{InducedIsomorphism}
 T_f\colon V_{\tilde{Y}} \to V_Y, \qquad \psi \longmapsto f^*\psi
\end{equation}
by pullback.
These assignments together constitute a contravariant symmetric monoidal functor
\begin{gather}
 \textsc{sMan}^{d-1}_{\mathrm{cl}} \longrightarrow \pHilb_\otimes^{d-1}\label{FunctorSManPHilb}
\end{gather}
from the category of closed $(d-1)$-dimensional spin$^{(c)}$-manifolds and spin isometries (with disjoint union as symmetric monoidal structure) to the category of polarized degree $d-1$ Hilbert spaces.
\end{Construction}

\begin{Remark}
The index of the polarized Hilbert space $V_Y = (V_Y, [L_Y])$ (see Definition~\ref{DefinitionIndexHilbertSpace}) associated to a spin manifold $Y$ can be written in terms of the index of the $\Cl_{d-1}$-linear index of the Dirac operator $D_Y$:
\begin{enumerate}\itemsep=0pt
\item[(a)]
In the spin case, where $V_Y$ is a real Hilbert space, this index is defined as the class of the graded $\Cl_d$-module $\ker(D_Y)$ in \[\mathfrak{M}_{d-1}/i^* \mathfrak{M}_{d} \cong KO^{-(d-1)}(*).\]
This class is by definition the alpha-invariant $\alpha(Y)$; see \cite[Section~4.2]{HitchinHarmonicSpinors}, \cite[Section~III.16]{LawsonMichelsohn}.
\item[(b)]
For a spin$^c$ manifold, the index of $V_Y$ coincides with the class of the complex graded $\Cl_{d-1}$-module $\ind(D_Y)$ in \[\mathfrak{M}_{d-1}^\C/i^* \mathfrak{M}_{d}^\C \cong KU^{-(d-1)}(*).\]
In this case, this group is either $0$ or $\Z$, depending on whether $d$ is even or odd, and $\ind(D_Y)$ can be identified with the ordinary index of the Dirac operator.
\end{enumerate}
\end{Remark}

The Dirac operator changes its sign when passing to opposites, $D_{Y^\vee} = - D_Y$, so that ${L_{Y^\vee} = \Gamma L_Y}$.
We therefore note the following.

\begin{Observation} \label{ObservationOppositeManifold}
The polarized Hilbert space $V_{Y^\vee}$ associated to the opposite spin manifold $Y^\vee$ of a closed $(d-1)$-dimensional closed spin manifold $Y$ is canonically isomorphic to the opposite polarized Hilbert space $\Pi V_Y$ (see Example~\ref{ExamplePolarizedHilbertSpace}\,(1)).
\end{Observation}

\subsection{Spin bordisms and harmonic spinors}

Let $X$ be a spin$^{(c)}$ manifold with boundary.
We say that $X$ has product structure near the boundary if the boundary has a neighborhood isometric to the metric product $\partial X \times [0, \varepsilon)$, for some $\varepsilon >0$.
We also require that the spinor connection is a product connection over this neighborhood.
Such a manifold has a well-defined \emph{double}, obtained by gluing two copies of $X$ (one with the reversed orientation and spin structure) together at the common boundary.
For details, see \cite[Section~I.9]{BBW}.

\begin{Definition}[bordisms]
Let $Y_0$ and $Y_1$ be two $(d-1)$-dimensional spin$^{(c)}$ manifolds.
\begin{enumerate}\itemsep=0pt
\item[(a)]
A \emph{bordism} from $Y_0$ to $Y_1$ is a $d$-dimensional compact spin$^{(c)}$ manifold-with-boundary $X$, with product structure near the boundary, together with a decomposition \[\partial X = \partial_1 X \sqcup \partial_0 X\] of the boundary and with spin isometries $f_1\colon Y_1 \to \partial_1 X$, $f_0\colon Y_0^\vee \to \partial_0 X$.
\item[(b)]
A \emph{thin bordism} between $Y_0$ and $Y_1$ is a $(d-1)$-dimensional spin$^{(c)}$ manifold $X$ together with spin isometries $f_1\colon Y_1 \to X$, $f_0\colon Y_0^\vee \to X^\vee$.
\item[(c)]
A \emph{bordism with thin parts} from $Y_0$ to $Y_1$ consists of decompositions $Y_i = Y_i^\prime \sqcup Y_i^{\prime\prime}$, an honest bordism $X^\prime\colon Y_0^\prime \to Y_1^\prime$ and a thin bordism $X^{\prime\prime}\colon Y_0^{\prime\prime} \to Y_1^{\prime\prime}$.
\end{enumerate}
\end{Definition}

If $Y_0$ and $Y_1$ are two $(d-1)$-dimensional spin- or spin$^{(c)}$ manifolds, there is a category \smash{$\sBord^{{(c)}}_d(Y_0, Y_1)$} whose objects are bordisms with thin parts from $Y_0$ to $Y_1$.
A morphism between two bordisms $(X, f_0, f_1)$, $\big(\tilde{X}, \tilde{f}_0, \tilde{f}_1\big)$ in this category is a spin isometry $h\colon X \to \tilde{X}$ such that
\begin{equation*}
 \tilde{f}_0 = h|_{\partial_0 X} \circ f_0 \qquad \text{and} \qquad \tilde{f}_1 = h|_{\partial_1 X} \circ f_1
\end{equation*}
This notion of morphism continues to make sense when $X$ has a thin part $X^{\prime\prime}$, provided that we write $\partial_0 X^{\prime\prime} = X^{\prime\prime}$ and $\partial_1 X^{\prime\prime} = (X^{\prime\prime})^\vee$.
Then $h$ is required to be an isomorphism of $d$-dimensional spin$^{(c)}$ manifolds on $X^\prime$ and an isomorphism of $(d-1)$-dimensional spin$^{(c)}$ manifolds on $X^{\prime\prime}$.
Of course, $\tilde{X}$ must have a thin part, too, in order for an isomorphism to exist, hand $h$ must map~$X^\prime$ to $\tilde{X}^\prime$ and $X^{\prime\prime}$ to $\tilde{X}^{\prime\prime}$.

If $X_{01} \colon Y_0 \to Y_1$ and $X_{12} \colon Y_1 \to Y_2$ are two bordisms, we may glue these bordisms together using the product structure near the middle boundary, obtaining a bordism $X_{12} \sqcup_{Y_1} X_{01}$ from~$Y_0$ to $Y_2$.
This gluing procedure extends in a straightforward way to bordisms with thin parts and provides composition functors
\begin{equation*}
\sBord^{{(c)}}_d(Y_1, Y_2) \times \sBord^{{(c)}}_d(Y_0, Y_1) \longrightarrow \sBord^{{(c)}}_d(Y_0, Y_2).
\end{equation*}

\begin{Definition}[Spin$^{(c)}$ bordism category]
The $d$-dimensional \emph{spin-bordism category} $\sBord_d$ is the bicategory with objects closed $(d-1)$-dimensional spin manifolds $Y$ and morphism categories the categories $\sBord_d(Y_0, Y_1)$ discussed above.
The $d$-dimensional \emph{spin$^c$-bordism category} $\sBord^c_d$ is defined analogously.
\end{Definition}

Disjoint union of manifolds endows $\sBord_d$ with a symmetric monoidal structure which extends the symmetric monoidal structure on the ordinary category $\textsc{sMan}^{d-1}_{\mathrm{cl}}$ of closed $(d-1)$-dimensional spin manifolds and spin isometries along the inclusion
\begin{equation}
\label{Spinclusion}
 \textsc{sMan}^{d-1}_{\mathrm{cl}} \longrightarrow \sBord_d
\end{equation}
(Explicitly, this functor is the identity on objects and sends a spin isometry $f\colon Y_0 \to Y_1$ to the thin bordism \smash{$Y_1 \stackrel{\id}{\to} Y_1 \stackrel{f}{\leftarrow} Y_0$}.)

We finish this section with the construction of a generalized Lagrangian $\mathcal{H}$ in $V_{\partial X}$ for each bordism $X$ between spin manifolds, which is needed for the construction of the desired field theory later on.
These results are essentially adaptations of the results of \cite[Section~3]{LudewigRoos} to our context.

Let $X$ be a compact spin manifold with boundary.
Consider the Dirac operator $D_X^+$ on $X$, which we view as unbounded operator on $L^2\big(X, \Sigma_X^+\big)$ with domain the smooth sections of $\Sigma_X^+$ that are supported in the interior of $X$.
The \emph{maximal Dirac operator}, denoted by $D_{\max}^+$, is by definition the adjoint of this operator.
It is a fact that there is a well-defined bijective boundary restriction map (the \emph{trace map})
\begin{equation*}
 \hat{r}\colon\ \dom\big(D_{\max}^+\big) \longrightarrow \check H(\partial X, \Sigma_{\partial X}), \qquad \Phi \longmapsto \Phi|_{\partial X}
\end{equation*}
taking values in a certain Sobolev type space of sections of $\Sigma_{\partial X}$, which in turn is contained in the Sobolev space of negative exponent $-1/2$.
For details, see \cite{BaerBallmannGuide}.
We denote by
\begin{equation*}
 \mathcal{H}_X := \big\{ \Phi \in \dom\big(D_{\max}^+\big) \mid D_X \Phi = 0, \Phi|_{\partial X} \in V_{\partial X}\big\},
\end{equation*}
the space of harmonic positive chirality spinors whose boundary restriction happens to be in the subspace $V_{\partial X} = L^2(\partial X, \Sigma_{\partial X})$.
The Clifford algebra $\Cl_{d-1}$ acts on $\mathcal{H}_X$ along the isomorphism~${\Cl_{d-1} \cong \Cl_d^+}$ from \eqref{IsomorphismCliffordDimensionShift} and the restriction map $r$ is $\Cl_{d-1}$-equivariant.
The image of $\mathcal{H}_X$ under the boundary restriction map $\hat{r}$ coincides with the $L^2$-closure
\begin{equation*}
 L_X = \overline{\big\{ \Psi|_{\partial X} \mid \Psi \in C^\infty\big(X, \Sigma_X^+\big) \colon D_X^+ \Psi = 0\big\}} \subseteq V_{\partial X}
\end{equation*}
of the space of boundary values of \emph{smooth} harmonic spinors.
We denote by $r$ the restriction of~$\hat{r}$ to $\mathcal{H}_X \subset \dom\big(D_{\max}^+\big)$.

\begin{Theorem} \label{ThmLXLagrangian}
If $X$ has product structure near the boundary, then $\mathcal{H}_X$ is a generalized Lagrangian in $V_{\partial X}$.
\end{Theorem}

\begin{proof}
That the kernel of $r$ is finite-dimensional follows from the \emph{unique continuation property} for Dirac operators \cite[Section~8]{BBW}: If $r(\Phi) = 0$, then $\Phi$ must vanish on the entire connected component containing $\partial X$.
We obtain that $ \ker(r) = \mathcal{H}_{X_{\mathrm{cl}}}$,
where $X_{\mathrm{cl}}$ denotes the union of all closed components of $X$.
But the space of harmonic spinors on a finite-dimensional manifold is always zero.

We must now show that $L_X$ is a Lagrangian in $V_{\partial X}$.
That $L_X$ is a $B$-isotropic follows from the integration by parts formula \cite[Theorem~3.2\,(5)]{BaerBallmannGuide}
\begin{equation*}
 \langle \Phi, D_X \Psi \rangle - \langle D_X \Phi, \Psi\rangle = \langle \nu \cdot \varphi,\psi\rangle, \qquad \Phi|_{\partial X} = \varphi,\qquad \Psi|_{\partial X} = \psi.
\end{equation*}
Indeed, when $\Phi, \Psi \in \mathcal{H}_X$, then also $D_X(\Phi \cdot e_d) \in \mathcal{H}_X$ (as the Dirac operator is $\Cl_d$-linear), hence
\begin{align*}
B(\varphi, \psi) = \langle \Gamma \varphi, \psi\rangle = \langle \nu \cdot \varphi \cdot e_d, \psi\rangle = \langle \Phi \cdot e_d, D_X \Psi \rangle - \langle D_X \Phi\cdot e_d, \Psi\rangle= 0.
\end{align*}
Showing that $L_X$ is in fact a Lagrangian is much more involved.
Here one considers the double~${X^\vee \sqcup_{\partial X} X}$, obtained by gluing $X$ together with its opposite $X^\vee$ along the common boundary, which is smooth as we assumed product structure near the boundary (for details on this construction, see the appendix of \cite{AmmannDahlHumbert} or \cite[Section~9]{BBW}).
Then the space of smooth sections of the spinor bundle $\Sigma_{\partial X}$ is the direct sum of the space of boundary values for harmonic spinors on $X$ and the corresponding space for $X^\vee$ \cite[Lemma 12.3]{BBW}.
Since the projection operator onto these subspaces are pseudodifferential operators of order zero \cite[Theorem~12.4]{BBW}, this result continues to hold for the $L^2$-closures of these subspaces.
In other words, we obtain that $V_{\partial X}$ is the direct sum of $L_X$ and $L_{X^\vee}$.

This latter statement is true in much more generality (see \cite[Section~XVII, Lemma B]{PalaisAtiyahSinger}), but in this special case, we can use the reflection symmetry of the double to identify $L_{X^\vee}$ with~$\Gamma L_X$ and to obtain that the decomposition $V_{\partial X} = L_X \oplus L_{X^\vee} = L_X \oplus \Gamma L_X$ is in fact orthogonal (see \cite[Theorem~3.12]{LudewigRoos}).
\end{proof}

\begin{Theorem}
\label{ThmClosed}
 $L_X$ is close to the sub-Lagrangian $L_{\partial X}$, defined in \eqref{APSLagrangian}.
\end{Theorem}

\begin{proof}
This is done by showing that both projections $P_{L_X}$ and $P_{L_{\partial X}}$ are pseudodifferential operators (of order zero), and that their principal symbols coincide.
Moreover, the product structure near the boundary yields that the difference is in fact a smoothing operator.
For details, see \cite[Remark~3.16]{LudewigRoos}.
\end{proof}

\begin{Remark}
It is here where it is important to choose the correct sign in \eqref{APSLagrangian}.
If we chose $L_{\partial X}$ to be the \emph{negative} spectral subspace of $D_{\partial X}$, then it would be close to the opposite Lagrangian $\Gamma L_X = L_X^\perp$.
\end{Remark}

For the generalized Lagrangians $\mathcal{H}_X$, we have the following gluing result.

\begin{Theorem} \label{TheoremFunctoriality}
Let $X$ be a compact spin manifold with boundary and let $Y \subseteq X$ be a compact hypersurface that separates $X$ into two parts $X_0$ and $X_1$.
Suppose that $X$ has product structure near the boundary.
Denote by $\mathcal{H}_{X_1} \circ \mathcal{H}_{X_0}$ the fiber product of $\mathcal{H}_{X_1}$ and $\mathcal{H}_{X_0}$ with respect to the boundary restriction maps $\mathcal{H}_{X_i} \to V_Y$.
Then the canonical map
\begin{equation*}
 \mathcal{H}_X \longrightarrow \mathcal{H}_{X_1} \circ \mathcal{H}_{X_0}, \qquad \Phi \longmapsto (\Phi|_{X_{1}}, \Phi|_{X_{0}})
\end{equation*}
is an isomorphism.
\end{Theorem}

\begin{proof}
It is clear that the map is injective.
To see surjectivity, let $(\Phi_{1}, \Phi_{0})$ be an element of the fiber product $\mathcal{H}_{X_{1}} \circ \mathcal{H}_{X_{0}}$, in other words, $\Phi_{i}$ are harmonic spinors which agree on $Y$.
Defining~$\Phi$~ by
\begin{equation*}
 \Phi(x) = \begin{cases} \Phi_{0}(x), & x \in X_{0}, \\ \Phi_{1}(x), & x \in X_{1}, \end{cases}
\end{equation*}
we obtain a continuous spinor on $X$ such that $D_X \Phi = 0$ in the interiors of $X_{0}$ and $X_{1}$.
But such a spinor is a weak solution to $D_X \Phi = 0$ on all of $X$ and must therefore by smooth by elliptic regularity; see \cite[Lemma 3.15]{LudewigRoos}.
\end{proof}

\subsection{The field theory} \label{SectionFermionicFieldTheory}

In this section, we construct the desired field theories
\begin{align*}
 \mathcal{L} \colon\ \sBord_d &\longrightarrow \LagCor^{d-1}_{\otimes, \R},\qquad
 \mathcal{L}^c \colon\ \sBord_d^c \longrightarrow \LagCor^{d-1}_{\otimes, \C}.
\end{align*}
For notational simplicity, we focus on the construction of $\mathcal{L}$.
The construction of $\mathcal{L}^c$ is entirely analogous.

Notice that $\mathcal{L}$ is a functor between bicategories.
These functors are often called pseudofunctors, alluding to the fact that they are not on the nose functorial, but only up to coherent natural transformations.
Explicitly, such a functor consists of
\begin{enumerate}\itemsep=0pt
\item[(1)]
For each object $Y$ of $\sBord_d$ an object $\mathcal{L}(Y)$ of $\LagCor_{\otimes, \R}^{d-1}$.
\item[(2)]
For any pair of objects $Y_0$, $Y_1$ of $\sBord_d$ a functor
\begin{equation}
\label{Lfunctor}
\mathcal{L}_{Y_0, Y_1}\colon\ \sBord_d(Y_0, Y_1) \longrightarrow \LagCor_{\otimes, \R}^{d-1}(\mathcal{L}(Y_0), \mathcal{L}(Y_1)).
\end{equation}
\item[(3)]
For each triple $Y_0$, $Y_1$, $Y_2$ of objects in $\sBord_d$ a natural transformation
\begin{equation}
\label{LambdaTransformation}
 \lambda_{Y_0, Y_1, Y_2}\colon\ \mathcal{L}_{Y_0, Y_2} \circ \mu \Longrightarrow \tilde{\mu} \circ (\mathcal{L}_{Y_1, Y_2} \times \mathcal{L}_{Y_0, Y_1}),
\end{equation}
where $\mu$ and $\tilde{\mu}$ denote the composition functors of $\sBord_d$, respectively $\LagCor_{\otimes, \R}^{d-1}$.
\end{enumerate}

A general pseudofunctor also requires the data of a 2-isomorphism
\begin{equation*}
\varepsilon_Y\colon\ \mathcal{L}_{Y, Y}(\id_Y) \Longrightarrow \id_{\mathcal{L}(Y)}
\end{equation*}
for each object $Y$ of $\sBord_d$.
However, $\mathcal{L}_{Y, Y}$ will send the identity of $Y$ to the identity of $\mathcal{L}(Y)$ on the nose, so these 2-morphisms will be identities in our case.
This implies in particular that the only non-trivial coherence condition for the data above is the commutativity of the diagram%
\begin{equation*}
\begin{tikzcd}[column sep=2.8cm]
\mathcal{L}_{03} \circ \mu \circ (\mu \times \id) \ar[r, "\mathcal{L}_{03} \circ \alpha", Rightarrow]
\ar[d, Rightarrow, "\lambda_{013} \circ (\mu \times \id)"']
	&
	\mathcal{L}_{03} \circ \mu \circ (\id \times \mu)
	\ar[d, Rightarrow, "\lambda_{023} \circ (\id \times \mu)"]
	\\
\tilde{\mu} \circ (\mathcal{L}_{13} \times \mathcal{L}_{01}) \circ (\mu \times \id)
	\ar[d, equal]
	&
	\tilde{\mu} \circ (\mathcal{L}_{23} \times \mathcal{L}_{02}) \circ ( \id \times \mu )
	\ar[d, equal]
\\
\tilde{\mu} \circ ((\mathcal{L}_{13} \circ \mu)\times \mathcal{L}_{01})
	\ar[d, Rightarrow, "\tilde{\mu} \circ (\lambda_{123} \times \mathcal{L}_{01})"']
	&
	\tilde{\mu} \circ (\mathcal{L}_{23} \times (\mathcal{L}_{02} \circ \mu ))
	\ar[d, Rightarrow, "\tilde{\mu} \circ (\mathcal{L}_{23} \times \lambda_{012})"]
\\
 \tilde{\mu} \circ ((\tilde{\mu} \circ (\mathcal{L}_{23} \times \mathcal{L}_{12})) \times \mathcal{L}_{01})
 		\ar[d, equal]
 		&
		\tilde{\mu} \circ (\mathcal{L}_{23} \times (\tilde{\mu} \circ ( \mathcal{L}_{12} \times \mathcal{L}_{01})))
		\ar[d, equal]
		\\
 \tilde{\mu} \circ (\tilde{\mu} \times \id) \circ (\mathcal{L}_{23} \times \mathcal{L}_{12} \times \mathcal{L}_{01})
 	\ar[r, Rightarrow, "\tilde{\alpha} \circ (\mathcal{L}_{23} \times \mathcal{L}_{12} \times \mathcal{L}_{01})"]
 		&
		\tilde{\mu} \circ (\id \times \tilde{\mu}) \circ (\mathcal{L}_{23} \times \mathcal{L}_{12} \times \mathcal{L}_{01}),
\end{tikzcd}
\end{equation*}
where $\alpha$ and $\tilde{\alpha}$ are the associators of $\sBord_d$, respectively $\LagCor_{\otimes, \R}^{d-1}$ and we abbreviated~$\mathcal{L}_{Y_i, Y_j}$ to $\mathcal{L}_{ij}$ and $\lambda_{Y_i, Y_j, Y_k}$ to $\lambda_{ijk}$.

We now describe the data (1)--(3) of the functor $\mathcal{L}$.

(1) If $Y$ is a $(d-1)$-dimensional spin manifold we set
\begin{equation*}
\mathcal{L}(Y) := V_Y = V_{Y_1} \oplus \cdots \oplus V_{Y_n},
\end{equation*}
where $V_Y$ is the polarized degree $d-1$ Hilbert space from Construction~\ref{ConstructionVX}.

(2)
We now define the functors $\mathcal{L}_{Y_0, Y_1}$ from \eqref{Lfunctor}, where $Y_0$ and $Y_1$ are two objects of $\sBord_d$.
For a bordism $(X, f_0, f_1)$ from $Y_0$ to $Y_1$, the boundary decomposes as $\partial X = \partial_0 X \sqcup \partial_1 X$, hence the corresponding Hilbert space is the direct sum $V_{\partial X} = V_{\partial_0 X} \oplus V_{\partial_1 X}$, using \eqref{ObservationMonoidality}.
Clearly, the Atiyah--Patodi--Singer sub-Lagrangian \eqref{APSLagrangian} decomposes correspondingly as
\begin{equation*}
 L_{\partial X} = L_{\partial_0 X} \oplus L_{\partial_1 X},
\end{equation*}
so it follows from Theorem~\ref{ThmClosed} that the generalized Lagrangian $\mathcal{H}_X$ is a split in $V_{\partial_0 X} \oplus V_{\partial_1 X}$.
Using Observation~\ref{ObservationOppositeManifold}, we moreover have an isomorphism
\begin{equation*}
\begin{tikzcd}[column sep=2cm]
V_{\partial X} \cong V_{\partial_0 X} \oplus V_{\partial_1 X}
\ar[r, "T_{f_0} \oplus T_{f_1}"] &
V_{Y_0^\vee} \oplus V_{Y_1} \cong \Pi V_{Y_0} \oplus V_{Y_1},
\end{tikzcd}
\end{equation*}
where $T_{f_0}\colon V_{\partial_0 X} \to V_{Y_0^\vee}$ and $T_{f_1}\colon V_{\partial_1 X} \to V_{Y_1}$ are the isomorphisms \eqref{InducedIsomorphism} of polarized degree~${d-1}$ Hilbert spaces induced by the spin isometries $f_i$ via pullback.
We now set
\begin{equation}
\label{LonThickBordism}
 \mathcal{L}_{Y_0, Y_1}(X, f_0, f_1) = \graph(T_{f_1}) \circ \mathcal{H}_X \circ \graph(T_{f_0}^*),
\end{equation}
which is a split generalized Lagrangian in $\Pi V_{Y_0} \oplus V_{Y_1}$ by Proposition~\ref{PropCompositionWithGraph}.

This construction has a straightforward generalization for bordism with thin parts.
First, if~$(X^{\prime\prime}, f_0, f_1)$ is thin, we just set
\begin{equation}
\label{LonThinBordism}
 \mathcal{L}_{Y_0, Y_1}(X^{\prime\prime}, f_0, f_1) = \graph(T_{f_1}) \circ \graph(T_{f_0}^*) = \graph(T_{f_0^{-1} \circ f_1}).
\end{equation}
This is an allowed morphism in $\LagCor_{\otimes, \R}^{d-1}$ as the spin isometry $f_0^{-1} \circ f_1$ clearly sends connected components of $Y_1$ to connected components of $Y_0$.
If now $(X = X^\prime \sqcup X^{\prime\prime}, f_0, f_1)$ is a morphisms with both thick and thin parts, we use \eqref{LonThickBordism} on the tick part $X^\prime$ and \eqref{LonThinBordism} on the thin part $X^{\prime\prime}$.

We have now defined the functor $\mathcal{L}_{Y_0, Y_1}$ on objects.
If $X, \tilde{X}\colon Y_0 \to Y_1$ are two bordisms and~${h\colon X \to \tilde{X}}$ is an isomorphism in the category $\sBord_d$, then $h$ induces an isomorphism of spans
\[
\mathcal{L}_{Y_0, Y_1}(X, f_0, f_1) \longrightarrow \mathcal{L}_{Y_0, Y_1}\big(\tilde{X}, \tilde{f}_0, \tilde{f}_1\big),
\]
 which is the identity on the thin part (observe here that
\begin{equation*}
f_0^{-1} \circ f_1 = (h \circ f_0)^{-1} \circ h \circ f_1 = \tilde{f}_0^{-1} \circ \tilde{f}_1
\end{equation*}
on the thin part $X^{\prime\prime}$), and on the thick part $X^{\prime}$ is induced by the vector space isomorphism $\mathcal{H}_{{X}^{\prime}} \to \mathcal{H}_{\tilde{X}^\prime}$ given by pullback with $h^{-1}$.

Observe that as claimed above, for each $Y$, $\mathcal{L}_{Y, Y}$ sends the identity morphism (which is the thin bordism given by the identities $Y \to Y$ and $Y^\vee \to Y^\vee$) to the identity correspondence in~$\LagCor_{\otimes, \R}^{d-1}(V_Y, V_Y)$.

(3)
For each triple of objects $Y_0$, $Y_1$, $Y_2$, the natural transformation $\lambda_{Y_0, Y_1, Y_2}$ from \eqref{LambdaTransformation} is given as follows.
For each pair $(X_{01}, f_0, f_1)\colon Y_0 \to Y_1$ and $\big(X_{12}, \tilde{f}_1, f_2\big) \colon Y_1 \to Y_2$ of composable bordisms with composition $(X_{02}, f_0, f_2)$, we must provide 2-isomorphisms
\begin{equation*}
\lambda_{X_{01}, X_{12}} \colon\ \mathcal{L}_{Y_0, Y_2}(X_{02}, f_0, f_2) \longrightarrow \mathcal{L}_{Y_1, Y_2}\big(X_{12}, \tilde{f}_1, f_2\big) \circ \mathcal{L}_{Y_0, Y_1}(X_{01}, f_0, f_1)
\end{equation*}
of generalized Lagrangians, i.e., isomorphisms in the category $\LagCor_{\otimes, \R}^{d-1}(V_{Y_0}, V_{Y_2})$, which are functorial in $X_{01}$ and $X_{12}$.
Let $X_{ij} = X^\prime_{ij} \sqcup X^{\prime\prime}_{ij}$ be the decomposition of $X$ into thick and thin parts.
On the thin part $X_{02}^{\prime\prime}$, both sides coincide and we take $\lambda_{X_{01}, X_{12}}$ to be just the identity.
On the thick part, we take it to be the restriction of harmonic spinors to the respective halves, which is an isomorphism by Theorem~\ref{TheoremFunctoriality}.

Since $\lambda_{Y_0, Y_1, Y_2}$ is given either by the identity or by restriction of harmonic spinors to subsets, commutativity of the relevant coherence diagram is trivial.
This finishes the construction of the functor $\mathcal{L}$.

It is possible to equip $\mathcal{L}$ with the structure of a symmetric monoidal functor of bicategories, whose main ingredients are the canonical isomorphisms
\begin{equation*}
 \mathcal{L}(Y) \oplus \mathcal{L}\big(\tilde{Y}\big) \longrightarrow \mathcal{L}\big(Y \sqcup \tilde{Y}\big)
\end{equation*}
for $(d-1)$-dimensional spin manifolds $Y$ and $\tilde{Y}$ and the isomorphisms
\begin{equation*}
 \mathcal{H}_{X} \oplus \mathcal{H}_{\tilde{X}} \cong \mathcal{H}_{X \sqcup \tilde{X}}
\end{equation*}
for any pair of bordisms $X \colon Y_0 \to Y_1$, $\tilde{X} \colon \tilde{Y}_0 \to \tilde{Y}_1$, which are natural in $X$ and $\tilde{X}$.
The resulting symmetric monoidal functor extends the functor \eqref{FunctorSManPHilb} along the inclusions \eqref{InclusionPHilbTimes} and~\eqref{Spinclusion}.
Spelling out completely the structure and coherence conditions for this functor is a rather long and tedious task and will be omitted here.

\subsection*{Acknowledgements}
I thank Ulrich Bunke for helpful discussions.
It is a pleasure to thank Christian B\"ar for his continuing support over many years.
I also thank the anonymous referees for their suggestions that helped improve the paper.
The author is indebted to SFB 1085 ``Higher Invariants'' funded by the Deutsche Forschungsgemeinschaft for financial support.

\pdfbookmark[1]{References}{ref}
\LastPageEnding

\end{document}